\documentclass[traditabstract]{aa} 
\usepackage{graphicx}
\usepackage{natbib}
\usepackage[switch]{lineno}
\modulolinenumbers[5]

\usepackage{amsmath,amssymb,wasysym}

\newcommand{\hatcurLCrprstar}{\ensuremath{0.1317\pm0.0007}}            
\newcommand{\hatcurLCimp}{\ensuremath{0.214_{-0.070}^{+0.061}}}        
\newcommand{\hatcurLCzeta}{\ensuremath{26.48\pm0.06}}                  
\newcommand{\hatcurLCdur}{\ensuremath{0.0859\pm0.0004}}                
\newcommand{\hatcurLCingdur}{\ensuremath{0.0104\pm0.0004}}             
\newcommand{\hatcurLCP}{\ensuremath{1.354133\pm0.000001}}              
\newcommand{\hatcurLCPshort}{\ensuremath{1.3541}}                      
\newcommand{\hatcurLCT}{\ensuremath{2455954.58576\pm0.00008}}          
\newcommand{\hatcurLBiz}{\ensuremath{0.2861}}                          
\newcommand{\hatcurLBiiz}{\ensuremath{0.2873}}                         
\newcommand{\hatcurLBii}{\ensuremath{0.3617}}                          
\newcommand{\hatcurLBiii}{\ensuremath{0.2744}}                         
\newcommand{\hatcurLBig}{\ensuremath{0.7052}}                          
\newcommand{\hatcurLBiig}{\ensuremath{0.1168}}                         
\newcommand{\hatcurLBir}{\ensuremath{0.4756}}                          
\newcommand{\hatcurLBiir}{\ensuremath{0.2487}}                         
\newcommand{\hatcurISOm}{\ensuremath{0.88\pm0.04}}                     
\newcommand{\hatcurISOr}{\ensuremath{0.89\pm0.02}}                     
\newcommand{\hatcurRVK}{\ensuremath{272.2\pm30.5}}                     
\newcommand{\hatcurRVjitterA}{\ensuremath{74.0}}                       
\newcommand{\hatcurRVjitterB}{\ensuremath{44.0}}                       
\newcommand{\hatcurRVjitterC}{\ensuremath{193.0}}                      
\newcommand{\hatcurPPi}{\ensuremath{87.8\pm0.7}}                       
\newcommand{\hatcurPPlogg}{\ensuremath{3.42\pm0.05}}                   
\newcommand{\hatcurPPar}{\ensuremath{5.57_{-0.09}^{+0.06}}}            
\newcommand{\hatcurPParel}{\ensuremath{0.0230\pm0.0003}}               
\newcommand{\hatcurPPrho}{\ensuremath{1.15\pm0.15}}                    
\newcommand{\hatcurPPm}{\ensuremath{1.37\pm0.16}}                      
\newcommand{\hatcurPPmlong}{\ensuremath{1.369\pm0.158}}                
\newcommand{\hatcurPPr}{\ensuremath{1.14\pm0.03}}                      
\newcommand{\hatcurPPrlong}{\ensuremath{1.139\pm0.025}}                
\newcommand{\hatcurPPmrcorr}{\ensuremath{0.11}}                        
\newcommand{\hatcurPPteff}{\ensuremath{1567\pm30}}                     
\newcommand{\hatcurPPtheta}{\ensuremath{0.062\pm0.007}}                
\newcommand{\hatcurPPfluxavg}{\ensuremath{1.36\pm0.11}}                
\newcommand{\hatcurLCrprstareccen}{\ensuremath{0.1326\pm0.0008}}       
\newcommand{\hatcurLCimpeccen}{\ensuremath{0.265_{-0.075}^{+0.053}}}   
\newcommand{\hatcurLCzetaeccen}{\ensuremath{26.58\pm0.07}}             
\newcommand{\hatcurLCdureccen}{\ensuremath{0.0859\pm0.0004}}           
\newcommand{\hatcurLCingdureccen}{\ensuremath{0.0107\pm0.0004}}        
\newcommand{\hatcurLCPeccen}{\ensuremath{1.354133\pm0.000001}}         
\newcommand{\hatcurLCTeccen}{\ensuremath{2455951.87748\pm0.00009}}     
\newcommand{\hatcurLBizeccen}{\ensuremath{0.2861}}                     
\newcommand{\hatcurLBiizeccen}{\ensuremath{0.2873}}                    
\newcommand{\hatcurLBiieccen}{\ensuremath{0.3617}}                     
\newcommand{\hatcurLBiiieccen}{\ensuremath{0.2744}}                    
\newcommand{\hatcurLBigeccen}{\ensuremath{0.7052}}                     
\newcommand{\hatcurLBiigeccen}{\ensuremath{0.1168}}                    
\newcommand{\hatcurLBireccen}{\ensuremath{0.4756}}                     
\newcommand{\hatcurLBiireccen}{\ensuremath{0.2487}}                    
\newcommand{\hatcurRVKeccen}{\ensuremath{278.7\pm33.0}}                
\newcommand{\hatcurRVkeccen}{\ensuremath{-0.033\pm0.052}}              
\newcommand{\hatcurRVheccen}{\ensuremath{-0.023\pm0.060}}              
\newcommand{\hatcurRVjitterAeccen}{\ensuremath{74.0}}                  
\newcommand{\hatcurRVjitterBeccen}{\ensuremath{44.0}}                  
\newcommand{\hatcurRVjitterCeccen}{\ensuremath{193.0}}                 
\newcommand{\hatcurRVecceneccen}{\ensuremath{0.071\pm0.049}}           
\newcommand{\hatcurRVomegaeccen}{\ensuremath{216\pm77}}                
\newcommand{\hatcurPPieccen}{\ensuremath{87.4\pm0.7}}                  
\newcommand{\hatcurPPloggeccen}{\ensuremath{3.43\pm0.08}}              
\newcommand{\hatcurPPareccen}{\ensuremath{5.65\pm0.32}}                
\newcommand{\hatcurPPareleccen}{\ensuremath{0.0230\pm0.0003}}          
\newcommand{\hatcurPPrhoeccen}{\ensuremath{1.20\pm0.28}}               
\newcommand{\hatcurPPmlongeccen}{\ensuremath{1.397\pm0.171}}           
\newcommand{\hatcurPPrlongeccen}{\ensuremath{1.131\pm0.065}}           
\newcommand{\hatcurPPmrcorreccen}{\ensuremath{-0.26}}                  
\newcommand{\hatcurPPteffeccen}{\ensuremath{1554\pm57}}                
\newcommand{\hatcurPPthetaeccen}{\ensuremath{0.064\pm0.009}}           
\newcommand{\hatcurPPfluxavgeccen}{\ensuremath{1.32\pm0.20}}           
\newcommand{\hatcurLCrprstarout}{\ensuremath{0.1335\pm0.0010}}            
\newcommand{\hatcurLCimpout}{\ensuremath{0.271_{-0.074}^{+0.055}}}        
\newcommand{\hatcurLCzetaout}{\ensuremath{26.52\pm0.07}}                  
\newcommand{\hatcurLCdurout}{\ensuremath{0.0862\pm0.0004}}                
\newcommand{\hatcurLCingdurout}{\ensuremath{0.0109\pm0.0005}}             
\newcommand{\hatcurLCPout}{\ensuremath{1.354133\pm0.000001}}              
\newcommand{\hatcurLCTout}{\ensuremath{2455954.58576\pm0.00009}}          
\newcommand{\hatcurSMEiteffout}{\ensuremath{5227\pm95}}                   
\newcommand{\hatcurSMEizfehout}{\ensuremath{0.15\pm0.05}}                 
\newcommand{\hatcurSMEiloggout}{\ensuremath{4.44\pm0.12}}                 
\newcommand{\hatcurSMEivsinout}{\ensuremath{1.5\pm0.5}}                   
\newcommand{\hatcurLBizout}{\ensuremath{0.2861}}                          
\newcommand{\hatcurLBiizout}{\ensuremath{0.2873}}                         
\newcommand{\hatcurLBiiout}{\ensuremath{0.3617}}                          
\newcommand{\hatcurLBiiiout}{\ensuremath{0.2744}}                         
\newcommand{\hatcurLBigout}{\ensuremath{0.7052}}                          
\newcommand{\hatcurLBiigout}{\ensuremath{0.1168}}                         
\newcommand{\hatcurLBirout}{\ensuremath{0.4756}}                          
\newcommand{\hatcurLBiirout}{\ensuremath{0.2487}}                         
\newcommand{\hatcurISOmlongout}{\ensuremath{0.882\pm0.037}}               
\newcommand{\hatcurISOrlongout}{\ensuremath{0.898\pm0.019}}               
\newcommand{\hatcurISOloggout}{\ensuremath{4.48\pm0.02}}                  
\newcommand{\hatcurISOlumout}{\ensuremath{0.54\pm0.06}}                   
\newcommand{\hatcurISOmvout}{\ensuremath{5.61\pm0.13}}                    
\newcommand{\hatcurISOageout}{\ensuremath{9.7\pm2.9}}                     
\newcommand{\hatcurISOMKout}{\ensuremath{3.64\pm0.07}}                    
\newcommand{\hatcurRVKout}{\ensuremath{268.9\pm29.0}}                     
\newcommand{\hatcurRVjitterAout}{\ensuremath{74.0}}                       
\newcommand{\hatcurRVjitterBout}{\ensuremath{44.0}}                       
\newcommand{\hatcurRVjitterCout}{\ensuremath{193.0}}                      
\newcommand{\hatcurPPiout}{\ensuremath{87.2\pm0.7}}                       
\newcommand{\hatcurPPloggout}{\ensuremath{3.39\pm0.05}}                   
\newcommand{\hatcurPParout}{\ensuremath{5.50\pm0.09}}                     
\newcommand{\hatcurPParelout}{\ensuremath{0.0230\pm0.0003}}               
\newcommand{\hatcurPPrhoout}{\ensuremath{1.05\pm0.14}}                    
\newcommand{\hatcurPPmlongout}{\ensuremath{1.345\pm0.150}}                
\newcommand{\hatcurPPrlongout}{\ensuremath{1.168\pm0.030}}                
\newcommand{\hatcurPPmrcorrout}{\ensuremath{0.08}}                        
\newcommand{\hatcurPPteffout}{\ensuremath{1577\pm31}}                     
\newcommand{\hatcurPPthetaout}{\ensuremath{0.060\pm0.007}}                
\newcommand{\hatcurPPfluxavgout}{\ensuremath{1.40\pm0.11}}                
\newcommand{\hatcurXdistout}{\ensuremath{360\pm11}}                       

\begin{document}
   \title{HATS-2b: A transiting extrasolar planet orbiting a K-type star
    showing starspot activity}

\titlerunning{HATS-2b}

\author{M.~Mohler-Fischer\inst{1}
\and
L.~Mancini\inst{1}
\and
J.~D.~Hartman\inst{4,5}
\and
G.~\'{A}.~Bakos\inst{4,5}\fnmsep\thanks{Alfred P.~Sloan Research Fellow}$^{,}$\thanks{Packard Fellow}
\and
K.~Penev\inst{4,5}
\and
D.~Bayliss\inst{6}
\and
A.~Jord\'an\inst{7}
\and
Z.~Csubry\inst{4,5}
\and
G.~Zhou\inst{6}
\and
M.~Rabus\inst{7}
\and
N.~Nikolov\inst{2}
\and
R.~Brahm\inst{7}
\and
N.~Espinoza\inst{7}
\and
L.~A.~Buchhave\inst{8,9}
\and
B.~B\'eky\inst{5}
\and
V.~Suc\inst{7}
\and
B.~Cs\'ak\inst{1,11}
\and
T.~Henning\inst{1}
\and
D.~J.~Wright\inst{3}
\and
C.~G.~Tinney\inst{3}
\and
B.~C.~Addison\inst{3}
\and
B.~Schmidt\inst{6}
\and
R.~W.~Noyes\inst{5}
\and
I.~Papp\inst{10}
\and
J.~L\'az\'ar\inst{10}
\and
P.~S\'ari\inst{10}
\and
P.~Conroy\inst{6}
              }
\institute{Max Planck Institute for Astronomy, K\"onigstuhl 17, D-69117 Heidelberg, Germany\\
         \email{mohler@mpia.de}
         \and
         Astrophysics Group, School of Physics, University of Exeter, Stocker Road, EX4 4QL, Exeter, UK
         \and
         Exoplanetary Science at UNSW, School of Physics, University of New
         South Wales, 2052, Australia and \\
         Australian Centre for Astrobiology, University of New South Wales, 2052, Australia
         \and
         Department of Astrophysical Sciences, Princeton University, NJ 08544, USA
         \and
         Harvard-Smithonian Center for Astrophysics, Cambridge, MA, USA
         \and
         Research School of Astronomy and Astrophysics, Australian National
         University, Canberra, ACT 2611, Australia
         \and
         Departamento de Astronom\'ia y Astrof\'isica, Pontificia Universidad Cat\'olica de Chile, Av. Vicu\~{n}a Mackenna 4860, 7820436 Macul, Santiago, Chile
         \and
         Niels Bohr Institute, University of Copenhagen, DK-2100, Copenhagen, Denmark
         \and
         Centre for Star and Planet Formation, Natural History Museum of Denmark,
         University of Copenhagen, DK-1350 Copenhagen, Denmark
         \and
         Hungarian Astronomical Association, Budapest, Hungary
         \and
         ELTE Gothard--Lend\"ulet Research Group, Szombathely, Hungary
             }

   \date{Received 08 April 2013; Accepted 26 June 2013}

\abstract{We report the discovery of HATS-2b, the second transiting
extrasolar planet detected by the HATSouth survey.  HATS-2b is moving
on a circular orbit around a $V=13.6$~mag, K-type dwarf
star (GSC 6665-00236), at a separation of \hatcurPParel~AU and with a
period of \hatcurLCPshort~days.  The planetary parameters have been
robustly determined using a simultaneous fit of the HATSouth, MPG/ESO~2.2\,m/GROND,
Faulkes Telescope South/Spectral transit photometry and MPG/ESO~2.2\,m/FEROS, Euler~1.2\,m/CORALIE, AAT~3.9\,m/CYCLOPS radial-velocity
measurements.  HATS-2b has a mass of \hatcurPPm~$M_{J}$, a
radius of \hatcurPPr~$R_{J}$ and an equilibrium temperature of
\hatcurPPteff~K.  The host star has a mass of
\hatcurISOm~$M_{\astrosun}$, radius of \hatcurISOr~$R_{\astrosun}$ and
shows starspot activity.  We characterized the stellar activity by
analysing two photometric-follow-up transit light curves taken with the
GROND instrument, both obtained simultaneously in four optical bands
(covering the wavelength range of $3860-9520$~\AA).  The two light
curves contain anomalies compatible with starspots on the photosphere
of the host star along the same transit chord.}


\maketitle

\section{Introduction}
\label{sec:1}

The first detection of a planet orbiting a main-sequence star (51 Peg;
\citealp{mayor1995}) started a new era of astronomy and planetary
sciences.  In the years since, the focus on exoplanetary discovery has
steadily increased, resulting in more than 850 planets being detected in 677
planetary systems\footnote{exoplanet.eu, as at 2013, March 28}. 
Statistical implications of the exoplanet discoveries, based on
different detection methods, have also been presented
(e.g.~\citealp{mayor2011,howard2012,cassan2012,fressin2013}).  Most of these
planets have been detected by the transit and radial velocity (RV) techniques. The former detects the decrease in a host star's brightness due to the transit of a planet in front of it, while the latter measures the Doppler shift of host star light due to stellar motion around the star-planet barycenter. In the case
of transiting extrasolar planets, the powerful combination of
both methods permits a direct estimate of mass and radius of the
planetary companion and therefore of the planetary average density and
surface gravity.  Such information is of fundamental importance in
establishing the correct theoretical framework of planet formation and
evolution (e.g.~\citealp{liu2011,mordasini2012a,mordasini2012b}).

Thanks to the effectiveness of ground- and space-based transit surveys
like TrES \citep{alonso2004}, XO \citep{mccullough2005}, HATNet
(e.g.~\citealp{bakos2012a,hartman2012}), HATSouth \citep{penev2013},
WASP (e.g.~\citealp{hellier2012,smalley2012}), QES
\citep{alsubai2011,bryan2012}, KELT \citep{siverd2012}, COROT
(e.g.~\citealp{rouan2012,patzold 2013}) and Kepler \citep{borucki2011a,
borucki2011b, batalha2012},
one third of the transiting exoplanets known today were detected in the past 2 years. 
In some cases, extensive follow-up campaigns have been necessary to
determine the correct physical properties of several planetary systems
(e.g.~\citealp{southworth2011,barros2011,mancini2013a}), or have been
used to discover other planets by measuring transit time variations
(e.g.~\citealp{rabus2009b}, \citealp{steffen2013}).  With high-quality photometric
observations it is also possible to detect transit anomalies which are
connected with physical phenomena, such as star spots
\citep{pont2007,rabus2009,desert2011,tregloan2013}, gravity darkening
\citep{barnes2009,szabo2011}, stellar pulsations \citep{cameron2010},
tidal distortion \citep{li2010,leconte2011}, and the presence of
additional bodies (exomoons) \citep{kipping2009,tusnski2011}.

In this paper we report the detection of HATS-2b, the second confirmed
exoplanet found by the HATSouth transit survey.  HATSouth is the first
global network of robotic wide-field telescopes, located at three sites
in the Southern hemisphere: Las Campanas Observatory (Chile), Siding
Spring Observatory (Australia) and H.E.S.S. (High Energy Stereoscopic System)  site (Namibia).  We refer
the reader to \citet{bakos2012b}, where the HATSouth instruments and
operations are described in detail.  HATS-2b is orbiting a K-type dwarf
star and has characteristics similar to those of most hot-Jupiter
detected so far.  Photometric follow-up performed during two transits
of this planet clearly reveals anomalies in the corresponding light
curves, which are very likely related to the starspot activity of the
host star.
%
\section{Observations}
\label{sec:2}

%
\subsection{Photometry}
\label{sec:2.1}
\begin{figure}
\centering
\includegraphics[width=9.cm]{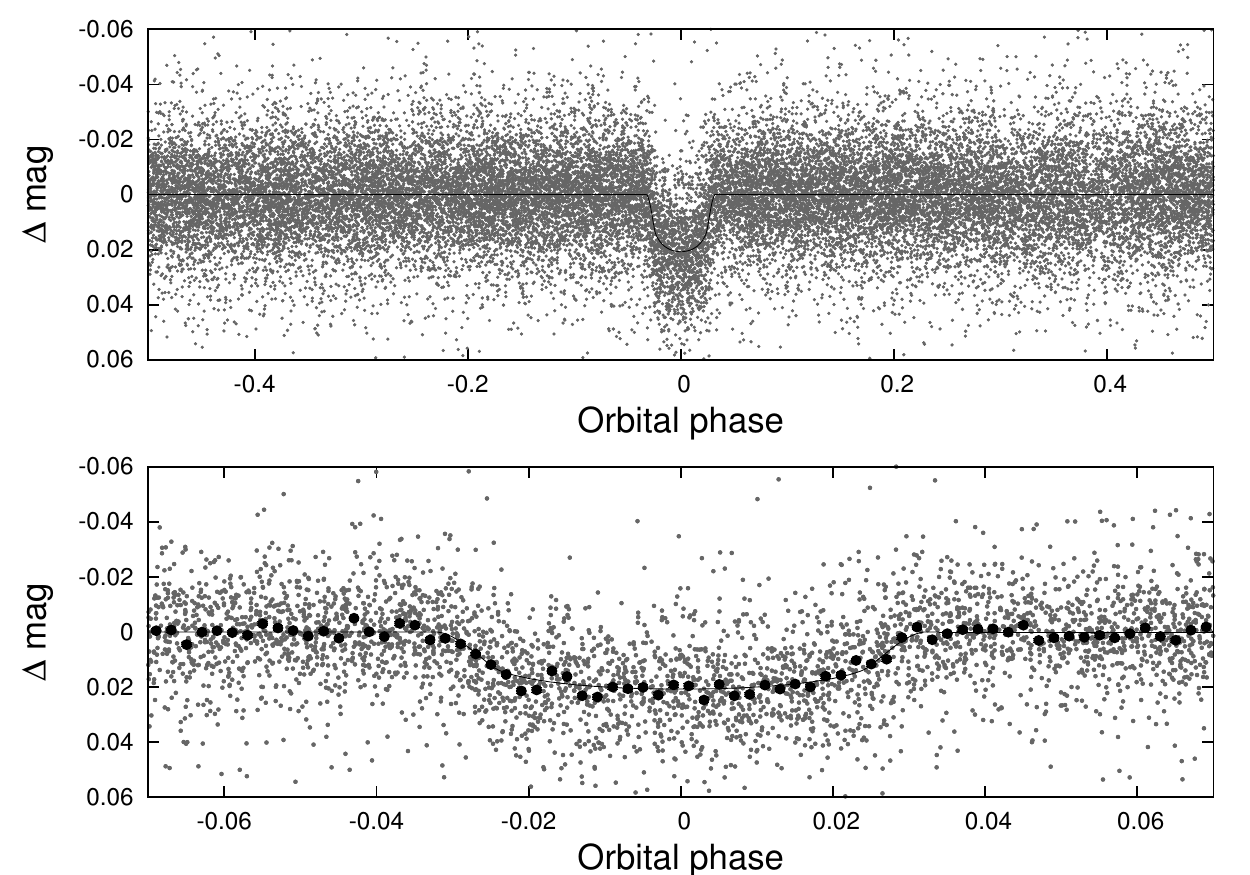}
\caption{\emph{Top panel:} 
Unbinned instrumental $r$-band light curve of HATS-2 folded with the
period P = 1.354133 days resulting from the global fit described in
Section \ref{sec:3}.  \emph{Lower panel:} zoom in on the transit; the
dark filled points show the light curve binned in phase using a bin
size of 0.002.  In both panels, the solid line shows the best-fit
transit model (see Section \ref{sec:phot_rv_analysis}).}
\label{fig:detection_lc}
\end{figure}

The star HATS-2 (GSC 6665-00236; $V=13.562 \pm 0.016$; J2000
$\alpha=11^{\mathrm{h}}46^{\mathrm{m}}57^{\mathrm{s}}.38$,
$\delta= -22^{\circ}33^{\prime}46^{\prime\prime}.77$, proper
motion $\mu_\alpha = -45.8 \pm 1.1$\,mas/yr, $\mu_\delta = -1.3 \pm
1.5$\,mas/yr; UCAC4 catalogue, \citealp{zacharias2012}) was identified as a
potential exoplanet host based on photometry from all the instruments
of the HATSouth facility (HS-1 to HS-6) between Jan 19 and Aug 10,
2010 (details are reported in Table~\ref{tab:phot_obs}).  The detection
light curve is shown in Figure~\ref{fig:detection_lc}.  This figure
shows that the discovery data is of sufficient quality that it permits fitting a \citet{mandel:2002} limb-darkened transit model.  A detailed overview
of the observations, the data reduction and analysis is given in
\citet{bakos2012b}. 
%

\begin{figure}[!h]
\centering
\includegraphics[trim = 15mm 18mm 65mm 85mm, clip,width=9.cm]{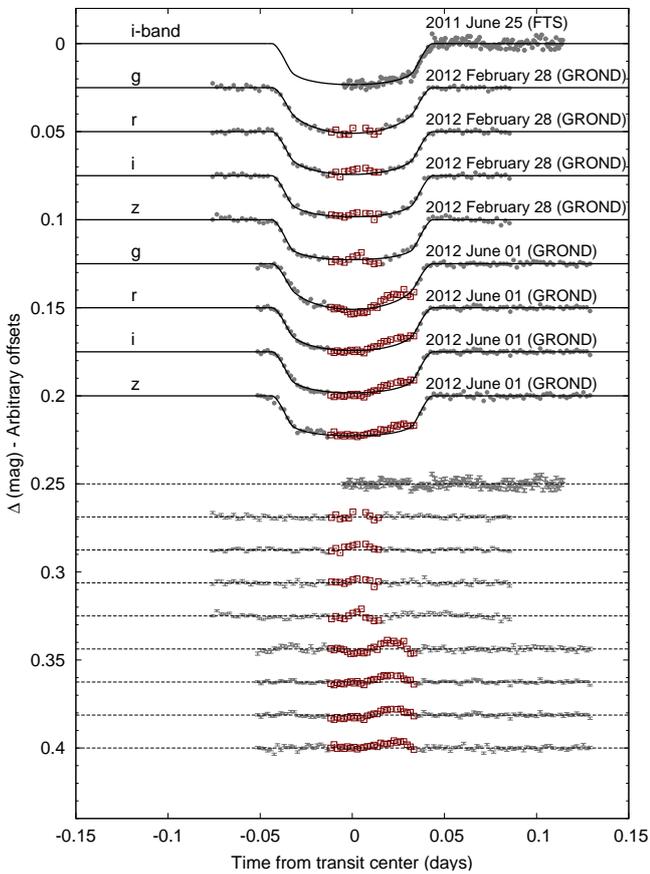}
\caption{
Phased transit light curves of HATS-2. The dates and instruments used
for each event are indicated.  The light curves are ordered according
to the date and to the central wavelength of the filter used (Sloan
$g$, $r$, $i$ and $z$).  Our best fit is shown by the solid lines (see
Section \ref{sec:phot_rv_analysis}).  Residuals from the fits are
displayed at the bottom, in the same order as the top curves.  The
GROND datapoints affected by anomalies are marked with red empty
squares and were not considered in estimating the final values of the
planetary-system parameters (see Sect.~\ref{sec:4}).
} %
\label{fig:phot_follow_up}
\end{figure}

HATS-2 was afterwards photometrically followed-up three times by two
different instruments at two different telescopes. On UT 2011, June 25, the mid-transit and the egress were observed 
with the “Spectral” imaging camera, mounted at 2.0 m Faulkes Telescope 
South (FTS), situated at Siding Spring Observatory (SSO) and operated 
as part of the Las Cumbres Observatory Global Telescope (LCOGT) network.  
The Spectral camera hosts a 4K$\times$4K array of $0.15^{\prime\prime}$ pixels, 
which is readout with $2\times2$ binning.  We defocus the telescope to reduce the effect 
of imperfect flat-fielding and to allow for longer exposure times without saturating.  We
use an $i$-band filter and exposure times of 30\,s, which with a 20\,s readout time gives 50\,s cadence 
photometry.  The data is calibrated with the automated LCOGT
reduction pipeline, which includes flat-field correction and fitting an
astrometric solution.  Photometry is performed on the reduced images
using an automated pipeline based on aperture photometry with Source
Extractor \citep{bertin:1996}.  The partial transit observed is shown in Figure~\ref{fig:phot_follow_up}, and permitted a 
refinement of the transit depth and ephemeris.
The latter was
particularly important for the subsequent follow-up observations
performed with the MPG\footnote{Max Planck
Gesellschaft}/ESO\,2.2m telescope at the La Silla Observatory (LSO).  Two full transits were covered on
February 28 and June 1, 2012, using GROND (\textbf{G}amma-\textbf{R}ay
Burst \textbf{O}ptical/\textbf{N}ear-Infrared \textbf{D}etector), which
is an imaging camera capable of simultaneous photometric observations
in four optical (identical to Sloan $g$, $r$, $i$, $z$) passbands
\citep{greiner2008}.  The main characteristics of the cameras and
details of the data reduction are described in \citet{penev2013}.  The
GROND observations were performed with the telescope defocussed and using relatively
long exposure times (80-90\,s, 150-200\,s cadence).  This way minimises noise sources
(e.g. flat-fielding errors, atmospheric variation or scintillation,
variation in seeing, bad tracking and Poisson noise) and delivers high-precision photometry of
transit events \citep{alonso2008, southworth2009}.  The light curves and
their best-fitting models are shown in Fig.  \ref{fig:phot_follow_up}. 
Distortions in the GROND light curves are clearly visible, which
we ascribe to stellar activity.  These patterns are analysed in
detail in Section 4.  Table~\ref{tab:phot_obs} gives an overview of all
the photometric observations for HATS-2.
%
%
%
\begin{table*}
\centering \caption{Summary of photometric observations of HATS-2}
\label{tab:phot_obs}
\begin{tabular}{l l r r c}\hline\hline
Facility & Date(s) & \# of images & Cadence (s) & Filter       \\
\hline
\textbf{Discovery} & & & & \\
HS-1 (Chile) & 2010, Jan 24 - Aug 09 & 5913 & 280 & Sloan $r$           \\
HS-2 (Chile) & 2010, Feb 11 - Aug 10 & 10195 & 280 & Sloan $r$          \\
HS-3 (Namibia) & 2010, Feb 12 - Aug 10 & 1159 & 280 & Sloan $r$           \\
HS-4 (Namibia) & 2010, Jan 26 - Aug 10 & 8405 & 280 & Sloan $r$           \\
HS-5 (Australia) & 2010, Jan 19 - Aug 08 & 640  & 280 & Sloan $r$           \\
HS-6 (Australia) & 2010, Aug 06 & 8 & 280 & Sloan $r$                       \\
 & & & & \\
\textbf{Follow-up} & & & & \\
FTS/Spectral & 2011, June 25 & 158 & 50 & Sloan $i$             \\
MPG/ESO 2.2\,m / GROND & 2012, February 28 & 69 &80 & Sloan $g$ \\
MPG/ESO 2.2\,m / GROND & 2012, February 28 & 70 &80 & Sloan $r$ \\
MPG/ESO 2.2\,m / GROND & 2012, February 28 & 69 &80 & Sloan $i$ \\
MPG/ESO 2.2\,m / GROND & 2012, February 28 & 71 &80 & Sloan $z$ \\
MPG/ESO 2.2\,m / GROND & 2012, June 1 & 99 &80 & Sloan $g$      \\
MPG/ESO 2.2\,m / GROND & 2012, June 1 & 99 &80 & Sloan $r$      \\
MPG/ESO 2.2\,m / GROND & 2012, June 1 & 99 &80 & Sloan $i$      \\
MPG/ESO 2.2\,m / GROND & 2012, June 1 & 99 &80 & Sloan $z$      \\
\hline
\end{tabular}
\end{table*}

%
\subsection{Spectroscopy}
\label{sec:spectroscopy}
HATS-2 was spectroscopically followed-up between May 2011 and April
2012 by five different instruments at five individual telescopes.  The
follow-up observations started in May 2011 with high signal to noise
(S/N) medium resolution ($\lambda/\Delta\lambda$ = 7000) reconnaissance
observations performed at the ANU 2.3\,m telescope located at SSO, with
the image slicing integral field spectrograph WiFeS \citep{dopita2007}. 
The results showed no RV variation with amplitude greater than
2\,km\,s$^{-1}$; this excludes most false-positive scenarios involving
eclipsing binaries.  Furthermore, an initial determination of the
stellar atmospheric parameters was possible ($T_{eff,\star}=4800\pm300\,K$, $\log g_{\star}=4.4\pm0.3$), indicating that HATS-2 is a dwarf
star.  Within the same month, high precision RV follow-up observations
started with the fibre-fed echelle spectrograph CORALIE
\citep{queloz2000b} at the Swiss Leonard Euler 1.2\,m telescope at LSO,
followed by further high precision RV measurements obtained with the
fibre-fed optical echelle spectrograph FEROS \citep{kaufer1998} at the
MPG/ESO 2.2\,m telescope at LSO\@.  Using the spectral synthesis code
\textit{SME} (`\textbf{S}pectroscopy \textbf{M}ade \textbf{E}asy',
\citealp{valenti1996}) on the FEROS spectra, it was possible to
determine more accurate values for the stellar parameters (see Sect. 
\ref{sec:stellar_parameter}).  Further RV measurements were obtained with
the CYCLOPS fibre-based integral field unit, feeding the
cross-dispersed echelle spectrograph UCLES, mounted at the 3.9\,m
Anglo-Australian Telescope (AAT) at SSO, and with the fibre-fed echelle
spectrograph FIES at the 2.5\,m telescope at the Nordic Optical
Telescope in La Palma.  We refer to \citet{penev2013} for a more
detailed description of the observations, the data reduction and the
RV determination methods for each individual instrument that we
utilized.

\begin{figure}
\centering
\includegraphics[trim = 15mm 18mm 70mm 110mm, clip,width=9.cm]{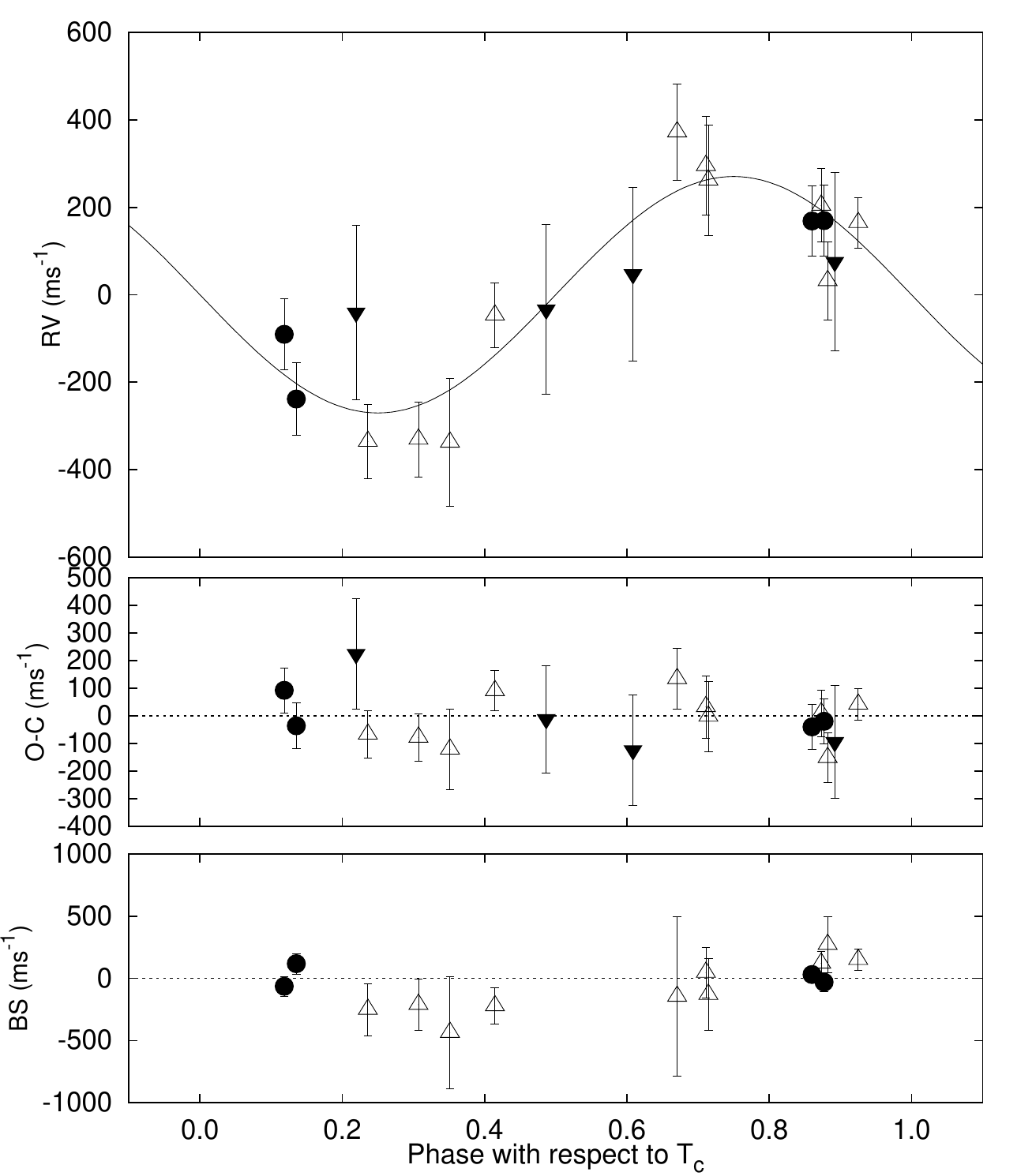}
\caption{
\emph{Top panel:} high-precision RV measurements for HATS-2 from
CORALIE (dark filled circles), FEROS (open triangles) and CYCLOPS
(filled triangles) shown as a function of orbital phase, together with
our best-fit model.  Zero phase corresponds to the time of mid-transit. 
The center-of-mass velocity has been subtracted.  \emph{Second panel:}
velocity O-C residuals from the best fit.  The error bars include a
component from astrophysical/instrumental jitter allowed to differ for
the three instruments (see Sec. 3.3).  \emph{Third panel:} bisector spans (BS), with
the mean value subtracted.  Note the different vertical scales of the
panels.
}
\label{fig:spec_follow_up}
\end{figure}

In total, 29 spectra were obtained, which are summarized in
Table~\ref{tab:spec_obs}.  Table~\ref{tab:RV_obs} provides the
high-precision RV and bisector span measurements. 
Figure~\ref{fig:spec_follow_up} shows the combined high-precision
RV measurements folded with the period of the transits. 
The error bars of the RV measurements include a component from
astrophysical/instrumental jitter allowed to differ for the three
instruments (Coralie: 74.0\,ms$^{-1}$, FEROS: 44.0\,ms$^{-1}$, CYCLOPS:
193.0\,ms$^{-1}$, see Sec. 3.3).

%
%
\begin{table*}
\centering %
\caption{Summary of spectroscopic observations of
HATS-2. Observing mode: HPRV = high-precision RV
measurements, RECON = reconnaissance observations.} %
\label{tab:spec_obs}
\begin{tabular}{l l r r r}\hline\hline
Telescope/Instrument & Date Range & \# of Observations &
Instrument resolution & Observing mode \\ \hline
ANU 2.3\,m/WiFeS & 2011, May 10-15 & 5 & 7000 & RECON\\
Euler 1.2\,m/Coralie & 2011, May 20-21 & 4 & 60000 & HPRV\\
ESO/MPG 2.2\,m/FEROS & 2011, June 09-25 & 9 & 48000 & HPRV\\
ESO/MPG 2.2\,m/FEROS & 2012, January 12 & 1 & 48000 & HPRV\\
ESO/MPG 2.2\,m/FEROS & 2012, March 04-06 & 2 & 48000 & HPRV\\
ESO/MPG 2.2\,m/FEROS & 2012, April 14-18 & 3 & 48000 & HPRV\\
AAT 3.9\,m/CYCLOPS & 2012, January 05-12 & 4 & 70000 & HPRV\\
NOT 2.5\,m/FIES & 2012, March 15 & 1 & 46000 & RECON\\ \hline
\end{tabular}
\end{table*}
%
%
\begin{table*}
\centering %
\caption{
Relative RV and bisector span (BS) measurements of HATS-2
from various instruments used for high-precision RV measurements
(c.f.~Table~\ref{tab:spec_obs}).  5 data points determined with FEROS
are not listed here and were not used for further analysis due to high
error bars caused by bad weather conditions.
}
\label{tab:RV_obs}
\begin{tabular}{l r r r r r r r r}\hline\hline
BJD & Relative RV & $\sigma_{\rm{RV}}$ & BS & $\sigma_{\rm{BS}}$ & Phase & Exp. Time & S/N & Instrument \\
(-2454000) & m\,s$^{-1}$) & (m\,s$^{-1}$) & (m\,s$^{-1}$) & (m\,s$^{-1}$) & & (s) & & \\
\hline
$ 1701.52346 $ & $   -90.37 $ & $    33.00 $ & $  -63.7 $ & $   79.5 $ & $   0.119 $ & 1800 & 9.0 & Coralie \\
$ 1701.54622 $ & $  -238.37 $ & $    36.00 $ & $  117.7 $ & $   83.3 $ & $   0.135 $ & 1800 & 8.0 & Coralie \\
$ 1702.52760 $ & $   168.63 $ & $    33.00 $ & $   31.3 $ & $   58.6 $ & $   0.860 $ & 1800 & 10.0 &Coralie \\
$ 1702.55065 $ & $   169.63 $ & $    33.00 $ & $  -30.8 $ & $   74.2 $ & $   0.877 $ & 1800 & 9.0 & Coralie \\
$ 1721.50300 $ & $   204.15 $ & $    71.76 $ & $  120.4 $ & $   93.8 $ & $   0.873 $ & 2400 & 14.0 &FEROS   \\
$ 1722.58300 $ & $   371.99 $ & $   100.77 $ & $ -146.9 $ & $  640.0 $ & $   0.671 $ & 2400 & 16.0 &FEROS   \\
$ 1723.44500 $ & $  -330.70 $ & $    73.66 $ & $ -211.0 $ & $  207.2 $ & $   0.307 $ & 2400 & 18.0 &FEROS   \\
$ 1736.46900 $ & $   164.63 $ & $    38.18 $ & $  149.3 $ & $   86.4 $ & $   0.925 $ & 2400 & 16.0 &FEROS   \\
$ 1737.53800 $ & $   261.69 $ & $   119.02 $ & $ -130.9 $ & $  287.5 $ & $   0.715 $ & 1044 & 17.0 &FEROS   \\
$ 1738.48600 $ & $   -47.00 $ & $    58.61 $ & $ -222.1 $ & $  146.9 $ & $   0.415 $ & 3000 & 12.0 &FEROS   \\
$ 1932.22448 $ & $   -33.63 $ & $    21.80 $ & $ 1114.3 $ & $   43.2 $ & $   0.487 $ & 2400 & 22.7 &CYCLOPS \\
$ 1933.21669 $ & $   -40.53 $ & $    51.20 $ & $ 3464.1 $ & $   19.9 $ & $   0.219 $ & 2400 & 20.7 &CYCLOPS \\
$ 1934.12774 $ & $    75.57 $ & $    65.58 $ & $ -4516.6$ & $  168.1 $ & $   0.892 $ & 2400 & 17.6 &CYCLOPS \\
$ 1938.81200 $ & $  -337.82 $ & $   139.13 $ & $ -436.4 $ & $  449.1 $ & $   0.351 $ & 2700 & 17.0 &FEROS   \\
$ 1939.16016 $ & $    47.07 $ & $    49.44 $ & $ 9189.5 $ & $ 1723.7 $ & $   0.608 $ & 2400 & 18.0 &CYCLOPS \\
$ 1990.75600 $ & $   294.99 $ & $   103.71 $ & $   46.3 $ & $  201.6 $ & $   0.711 $ & 2700 & 15.0 &FEROS   \\
$ 1992.82100 $ & $  -335.90 $ & $    72.26 $ & $ -251.5 $ & $  208.5 $ & $   0.236 $ & 2700 & 19.0 &FEROS   \\
$ 2035.67400 $ & $    31.55 $ & $    77.56 $ & $  270.6 $ & $  223.1 $ & $   0.882 $ & 3600 & 22.0 &FEROS   \\
\hline
\end{tabular}
\tablefoot{The Coralie RV uncertainties listed here are known to be
  underestimated. Updated estimates are available, but we list here the values
  we used in the analysis. We note that in any case a jitter is
  included in the analysis to account for any additional scatter to that
  implied by the uncertainties, see Sec. 3.3 (cf. Table \ref{tab:orbit_parameter}).
}
\end{table*}

%
\section{Analysis}
\label{sec:3}

%
\subsection{Stellar parameters}%
\label{sec:stellar_parameter}
As already mentioned in Sect.~\ref{sec:spectroscopy}, we estimated the
stellar parameters, i.e.~effective temperature $T_{eff\star}$,
metallicity [Fe/H], surface gravity $\log g$ and projected rotational
velocity $v \sin i$, by applying \textit{SME} on the high-resolution
FEROS spectra.  SME determines stellar and atomic parameters by fitting
spectra from model atmospheres to observed spectra and estimates the
parameter errors using the quality of the fit, expressed by the reduced
$\chi^2$, as indicator.  In case the S/N is not very high, or the
spectrum is contaminated with telluric absorption features, cosmics or
stellar emission lines, the reduced $\chi^2$ does not always converge
to unity, which leads to small errors for the stellar parameter values. 
To estimate of error bars, we used SME to determine the
stellar parameters of each FEROS spectrum and calculated the weighted mean and
corresponding scatter (weighted by the S/N of individual spectra). The results for the
spectroscopic stellar parameters including the assumed values for  micro- $v_{mic}$ and macroturbulence $v_{mac}$ of the SME analysis are listed in
Table~\ref{tab:stellar_parameter}.

By modeling the light curve alone it is possible to determine the
stellar mean density, which is closely related to the normalized
semimajor axis $a/R_{\star}$ (Sect.~\ref{sec:phot_rv_analysis})
assuming a circular orbit.  Furthermore, adding RV measurements allows
the determination of these parameters for elliptical orbits as well.

\begin{figure}
\centering
\includegraphics[width=9.0cm]{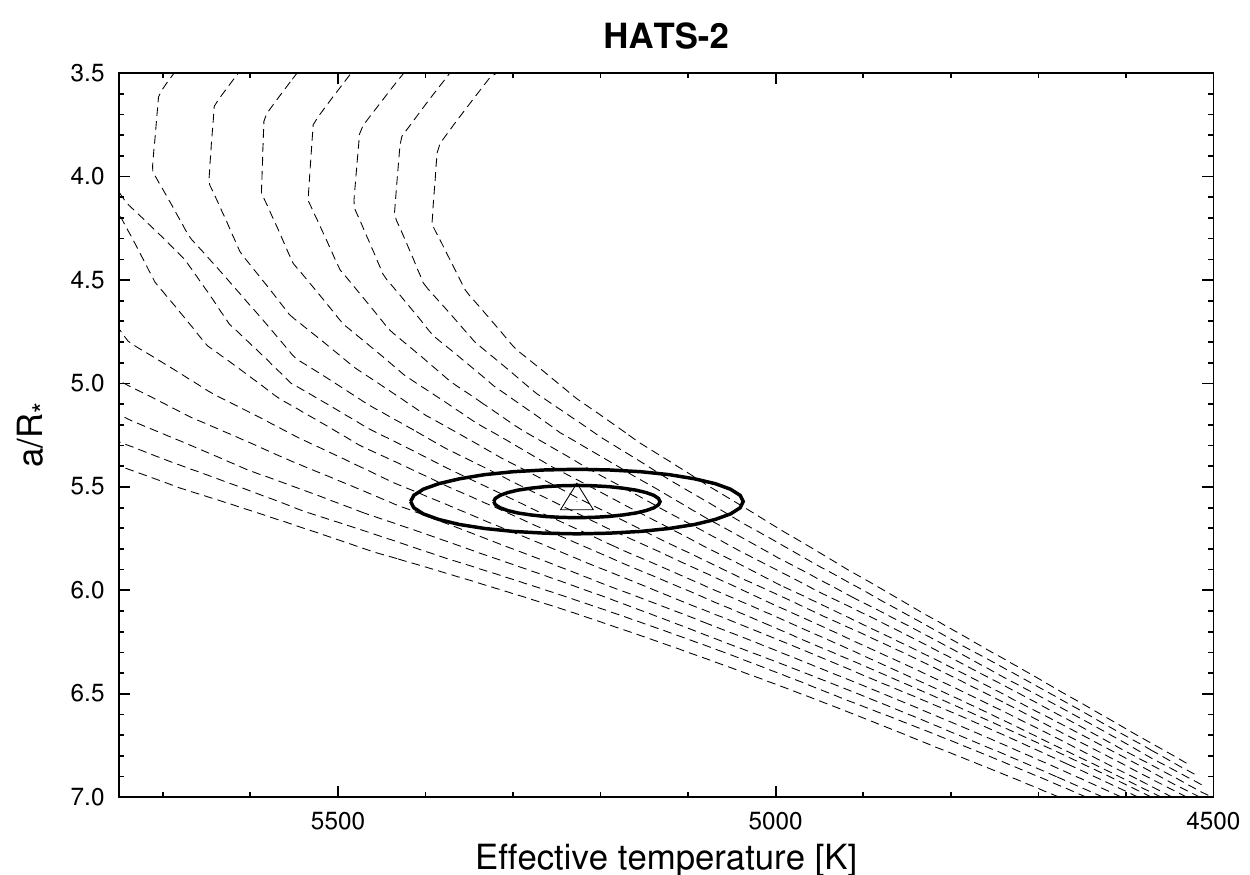}
\caption{
YY-isochrones from \citet{yi2001} for the metallicity of [Fe/H]$=0.15$. 
Isochrones are plotted for ages between 1 and 14\,Gyr in steps of
1\,Gyr (left to right).  The ellipses mark the 1-$\sigma$ and
2-$\sigma$ confidence ellipsoids for the determined values of
$a/R_{\star}$ and $T_{eff\star}$. The isochrones plotted here have a fixed metallicity for visualization purposes only, uncertainties on the metallicity are propagated into the uncertainties on the stellar mass and radius.
}
\label{fig:YY-diagram}
\end{figure}

To obtain the light curve model, quadratic limb-darkening coefficients
are needed, which were determined using \citet{claret2004} and the
initially determined stellar spectroscopic parameters.  We used the
Yonsei-Yale stellar evolution models (\citealp{yi2001}; hereafter YY)
to determine fundamental stellar parameters such as the mass, radius,
age and luminosity.  The light curve based stellar mean density and
spectroscopy based effective temperature and metallicity, coupled with isochrone
analysis, together permit a more accurate stellar surface gravity
determination. To allow uncertainties in the measured parameters to propagate into the stellar physical parameters we assign an effective temperature and metallicity, drawn from uncorrelated Gaussian distributions, to each stellar mean density in our MCMC chain, and perform the isochrone look-up for each link in the MCMC chain. The newly determined value for $\log g_{\star} = 4.50
\pm 0.05$ is consistent with the initial value of $\log g_{\star} =
4.44 \pm 0.12$ thus we refrain from re-analyzing the spectra fixing the
surface gravity to the revised value.
%
\begin{table}
\centering %
\tiny %
\caption{Stellar parameters for HATS-2}
\label{tab:stellar_parameter}
\begin{tabular}{l r l}%
\hline
\hline
Parameter & Value & Source \\ %
\hline
\textbf{Spectroscopic properties} & &                       \\
$T_{eff\star}$ (K)\dotfill          & \hatcurSMEiteffout & SME \\
$[Fe/H]$ \dotfill                   & \hatcurSMEizfehout & SME \\
$v \sin i_{\star}$ (km\,s$^{-1}$) \dotfill  & \hatcurSMEivsinout & SME \\
$\log g_{\star}$ (cgs) \dotfill     & \hatcurSMEiloggout & SME \\
$v_{mic}$ (km\,s$^{-1}$)$^a$ \dotfill & 1.5 & SME \\
$v_{mac}$ (km\,s$^{-1}$)$^a$ \dotfill & 2.0 & SME \\

                                  & &                       \\
\textbf{Photometric properties}   & &                       \\
$V$ (mag) \dotfill & $13.562 \pm 0.016$ & APASS$^1$              \\
$B$ (mag) \dotfill & $14.490 \pm 0.031$ & APASS              \\
$g$ (mag) \dotfill & $13.991 \pm 0.012$ & APASS              \\
$r$ (mag) \dotfill & $13.260 \pm 0.020$ & APASS              \\
$i$ (mag) \dotfill & $13.018 \pm 0.021$ & APASS              \\
$J$ (mag) \dotfill & $11.906 \pm 0.024$ & 2MASS$^2$              \\
$H$ (mag) \dotfill & $11.475 \pm 0.023$ & 2MASS              \\
$K$ (mag) \dotfill & $11.386 \pm 0.023$ & 2MASS              \\
                                  & &                        \\
\textbf{Derived properties}       & &                        \\
$M_{\star}$ ($M_{\astrosun}$) \dotfill  & \hatcurISOmlongout & YY+$a/R_\star$+SME \\
$R_{\star}$ ($R_{\astrosun}$) \dotfill  & \hatcurISOrlongout & YY+$a/R_\star$+SME \\
$\log g_{\star}$ (cgs)    \dotfill  & \hatcurISOloggout  & YY+$a/R_\star$+SME \\
$L_{\star}$ ($L_{\astrosun}$) \dotfill  & \hatcurISOlumout   & YY+$a/R_\star$+SME \\
$M_V$ (mag)                   \dotfill  & \hatcurISOmvout    & YY+$a/R_\star$+SME \\
$M_K$ (mag)                   \dotfill  & \hatcurISOMKout    & YY+$a/R_\star$+SME \\
Age (Gyr)                     \dotfill  & \hatcurISOageout   & YY+$a/R_\star$+SME \\
Distance (pc)$^b$             \dotfill  & \hatcurXdistout    & YY+$a/R_\star$+SME \\
\boldmath{$P_{\star,\rm{rot}}$} \textbf{(d)}$^c$ \dotfill & \boldmath{$30.32 \pm 10.13$} & \\
\hline
\end{tabular}
\tablefoot{
$^1$ The \textbf{A}AVSO (\textbf{A}merican \textbf{A}ssociation of
\textbf{V}ariable \textbf{S}tar \textbf{O}bservers)
\textbf{P}hotometric All-\textbf{S}ky \textbf{S}urvey\\
$^2$ \textbf{Two} \textbf{M}icron \textbf{A}ll \textbf{S}ky \textbf{S}urvey \\
$^a$ given values for micro- ($v_{mic}$) and macroturbulence ($v_{mac}$) are initial guesses, which were fixed during the analysis. Afterwards, the values were set free, but parameters were consistent with the fixed scenario within errorbars. Therefore, the stellar parameters given here and used throughout the following analysis are the ones determined with fixed micro- and macroturbulence\\
$^b$ $A_\mathrm{V}$ corrected\\
$\boldmath{^c}$ upper limit of the rotational period of HATS-2 using the determined values for $v \sin i_\star$ and $R_\star$.}

\end{table}
The spectroscopic, photometric and derived stellar properties are
listed in Table~\ref{tab:stellar_parameter}, whereas the adopted
quadratic limb-darkening coefficients for the individual photometric
filters are shown in Table~\ref{tab:orbit_parameter}.
To illustrate the position of HATS-2 in the H-R diagram, we
plotted the normalized semi-major axis $a/R_{\star}$ versus effective
temperature $T_{eff\star}$.  Figure~\ref{fig:YY-diagram} shows the
values for HATS-2 with their 1-$\sigma$ and 2-$\sigma$ confidence
ellipsoids as well as YY-isochrones calculated for the determined
metallicity of [Fe/H]$=0.15$ and interpolated to values between 1 and
14\,Gyr in 1\,Gyr increments from our adopted model. 
%

\subsection{Stellar rotation}
\begin{figure}
\centering
\includegraphics[trim= 20mm 15mm 65mm 190mm, clip,width=9.0cm]{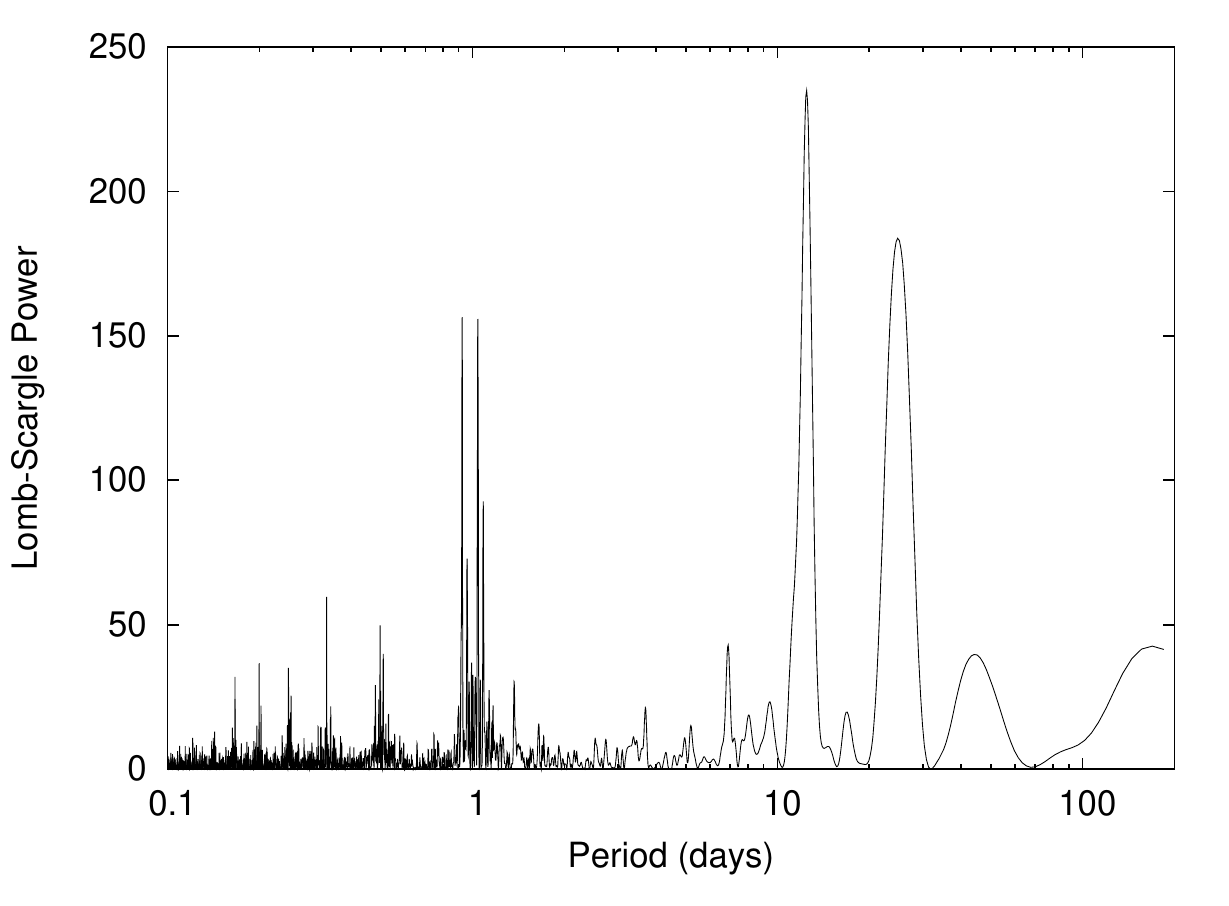}
\caption{Normalized Lomb-Scargle periodogram of the combined HATSouth light curve of HATS-2. Transits have been removed from the data before applying the periodogram. A strong signal with a period of 12.46 days is detected in the data.}
\label{fig:periodogram}
\end{figure}

We applied the Lomb-Scargle periodogram \citep{lomb:1976,scargle:1982} to the HATSouth light curve for HATS-2 and found a significant peak at a period of $P = 12.46 \pm 0.02$\,d with a S/N measured in the periodogram of $87$ and a formal false alarm probability of $10^{-98}$ \textbf{calculated following \cite{press:1992}}. Fig. \ref{fig:periodogram} shows the normalized Lomb-Scargle periodogram of the HATSouth light curve. The peak-to-peak amplitude of the signal over the full 203\,d spanned by the observations is 7.4\,mmag. If we split the data into bins of duration 50\,d, the amplitude in each bin varies from 3.6\,mmag to 10.0\,mmag. We interpret this signal as being due to starspots modulated by the rotation of the star. The stellar rotation period is thus $\sim 12.5$\,d, or twice this value (as seen in many open clusters, an individual star often shows two minima per cycle so that the rotation period is double the value found from a periodogram analysis; also note that due to differential rotation and the unknown latitudinal distribution of spots on the star, the equatorial period may be as much as 10--20\% shorter than the measured period). 
Both rotation periods (12.5 and 25\,d) are consistent with the upper limit of $P_{\star,\rm{rot}}$ of $30.32 \pm 10.13\,$d derived from the determined $v \sin i$ and $R_\star$ (see Tab. 4). 
The rotation period of $12.46$\,d is comparable to that of similar-size stars in the 1\,Gyr open cluster NGC~6811 \citep{meibom:2011}, which shows a tight period--color sequence. The spin-down rate for sub-solar-mass stars is poorly constrained beyond 1\,Gyr, but assuming a \citet{skumanich:1972} spin-down of $P \propto t^{0.5}$, the expected rotation period reaches $\sim 25$\,d at an age of $4$\,Gyr. Based on this we estimate a gyrochronology age of either $\sim 1$\,Gyr, or $\sim 4$\,Gyr for HATS-2, depending on the ambiguous rotation period.

%
\subsection{Excluding blend scenarios}
\label{sec:3.2}

To rule out the possibility that HATS-2 is actually a blended stellar
system mimicking a transiting planet system we conduct a detailed
modeling of the light curves following the procedure described in
\citet{hartman2011}.  Based on this analysis we can reject hierarchical
triple star systems with greater than $4.5\sigma$ confidence, and
blends between a foreground star and a background eclipsing binary with
$\sim 4\sigma$ confidence.  Moreover, the only non-planetary blend
scenarios which could plausibly fit the light curves (ones that cannot
be rejected with greater than $5\sigma$ confidence) are scenarios which
would have easily been rejected by the spectroscopic observations
(these would be obviously double-lined systems, also yielding several
km\,s$^{-1}$ RV and/or BS variations).  We thus conclude that the
observed transit is caused by a planetary companion orbiting HATS-2.

%
\subsection{Simultaneous analysis of photometry and radial velocity} %
\label{sec:phot_rv_analysis}
Following \citet{bakos2010} we correct for systematic noise in the follow-up light curves by applying external parameter decorrelation and the Trend Filtering Algorithm (TFA) simultaneously with our fit. For the FTS light curve we decorrelate against the hour angle of the observations (to second order), together with three parameters describing the profile shape (to first order), and we apply TFA. For the GROND light curves we only decorrelate against the hour angle as the PSF model adopted for FTS is not applicable to GROND, and the number of neighboring stars that could be used in TFA is small. Following the procedure described in \citet{bakos2010}, the FTS and
GROND photometric follow-up measurements (Table~\ref{tab:phot_obs})
were simultaneously fitted with the high-precision RV measurements
(Table~\ref{tab:RV_obs}) and HATSouth photometry.  The light curve parameters, RV parameters,
and planetary parameters are listed in Table~\ref{tab:orbit_parameter}.

Table~\ref{tab:orbit_parameter} also contains values for the radial
velocity jitter for all three instruments used for high-precision RV
measurements.  They are added in quadrature to the RV results of the
particular instrument.  These values are determined such that $\chi^2$
per degree of freedom equals unity for each instrument when fitting a
fiducial model.  If $\chi^2$ per degree of freedom is smaller than
unity for that instrument, then no jitter would be added.  The RV
jitters are empirical numbers that are added to the measurements such
that the actual scatter in the RV observations sets the posterior
distributions on parameters like the RV semi-amplitude.

Allowing the orbital eccentricity to vary during the simultaneous fit,
we include the uncertainty for this value in the other physical
parameters.  We find that the observations are consistent with a
circular orbit ($e = 0.071 \pm 0.049$) and we therefore fix the eccentricity
to zero for the rest of this analysis. Table~\ref{tab:orbit_parameter} shows that the derived parameters
obtained by including the distorted regions of the light curves are
consistent with those derived with these regions excluded,
indicating that the starspots themselves are not affecting the stellar
or planet parameters in a significant way. \\
The RMS varies from 1 to 1.6 mmag for the complete light curves and 0.8 to 1.3 mmag when then spot-affected regions are excluded, respectively. We scaled the photometric uncertainties for each of the light curves such that $\chi^2$ per degree of freedom equals one about the best-fit model.
We adopt the parameters obtained with the light curve
distortions excluded in a fixed circular orbit.

\begin{table*}
\centering %
\tiny %
\caption{
Orbital and planetary parameters for the HATS-2 system for different
fitting scenarios: including the light curve distortions with free and
fixed eccentricity $e$ as well as excluding the light curve distortions with a
fixed circular orbit.  The last scenario was adopted for further
analysis steps (parameters are highlighted in bold font).
}
\label{tab:orbit_parameter}
\begin{tabular}{l r r r}%
\hline \hline
Parameter & LC distortions included, $e\equiv0$ & LC distortions
included, free $e$ & \textbf{LC distortions excluded,
}\boldmath{$e\equiv0$}\\
\hline
\textbf{Light curve parameters} & \\
$P$ (days)                      \dotfill & \hatcurLCP    &  \hatcurLCPeccen &  \boldmath{\hatcurLCPout} \\
$T_c$ (BJD)$^a$                 \dotfill & \hatcurLCT    &   \hatcurLCTeccen & \boldmath{\hatcurLCTout} \\
$T_{14}$ (days)$^a$             \dotfill & \hatcurLCdur  &    \hatcurLCdureccen & \boldmath{\hatcurLCdurout} \\
$T_{12}$ = $T_{34}$ (days)$^a$  \dotfill & \hatcurLCingdur &  \hatcurLCingdureccen & \boldmath{\hatcurLCingdurout}\\
$a/R_{\star}$                   \dotfill & \hatcurPPar     &  \hatcurPPareccen & \boldmath{\hatcurPParout}\\
$\zeta/R_{\star}$$^b$           \dotfill & \hatcurLCzeta   &  \hatcurLCzetaeccen & \boldmath{\hatcurLCzetaout} \\
$R_p/R_{\star}$                 \dotfill & \hatcurLCrprstar & \hatcurLCrprstareccen & \boldmath{\hatcurLCrprstarout} \\
$b \equiv a \cos i_p / R_{\star}$ \dotfill & \hatcurLCimp    &  \hatcurLCimpeccen & \boldmath{\hatcurLCimpout}\\
$i_p$ (deg)                       \dotfill & \hatcurPPi    &   \hatcurPPieccen &  \boldmath{\hatcurPPiout}\\
& \\
\textbf{Limb-darkening coefficients}$^c$ & \\
$a_g$ (linear term)    \dotfill & \hatcurLBig  &  \hatcurLBigeccen & \boldmath{\hatcurLBigout}\\
$b_g$ (quadratic term) \dotfill & \hatcurLBiig  & \hatcurLBiigeccen & \boldmath{\hatcurLBiigout}\\
$a_r$                  \dotfill & \hatcurLBir   & \hatcurLBireccen & \boldmath{\hatcurLBirout}\\
$b_r$                  \dotfill & \hatcurLBiir  & \hatcurLBiireccen & \boldmath{\hatcurLBiirout}\\
$a_i$                  \dotfill & \hatcurLBii   & \hatcurLBiieccen & \boldmath{\hatcurLBiiout}\\
$b_i$                  \dotfill & \hatcurLBiii  & \hatcurLBiiieccen & \boldmath{\hatcurLBiiiout} \\
$a_z$                  \dotfill & \hatcurLBiz   & \hatcurLBizeccen & \boldmath{\hatcurLBizout}\\
$b_z$                  \dotfill & \hatcurLBiiz  & \hatcurLBiizeccen & \boldmath{\hatcurLBiizout}\\
& \\
\textbf{Radial velocity parameters} & \\
$K$ (m\,s$^{-1}$)               \dotfill & \hatcurRVK      & \hatcurRVKeccen & \boldmath{\hatcurRVKout} \\
$e \cos \omega$                 \dotfill & $0.000$     & \hatcurRVkeccen & \boldmath{$0.000$}  \\
$e \sin \omega$                \dotfill & $0.000$     & \hatcurRVheccen & \boldmath{$0.000$}  \\
$e$$^d$                       \dotfill &  $0.000$ & \hatcurRVecceneccen & \boldmath{$0.000$} \\
$\omega$ (deg)                  \dotfill & $0.000$  & \hatcurRVomegaeccen & \boldmath{$0.000$} \\
RV jitter Coralie (m\,s$^{-1}$) \dotfill & \hatcurRVjitterA & \hatcurRVjitterAeccen & \boldmath{\hatcurRVjitterAout}\\
RV jitter FEROS   (m\,s$^{-1}$) \dotfill & \hatcurRVjitterB & \hatcurRVjitterBeccen & \boldmath{\hatcurRVjitterBout}\\
RV jitter CYCLOPS (m\,s$^{-1}$) \dotfill & \hatcurRVjitterC & \hatcurRVjitterCeccen & \boldmath{\hatcurRVjitterCout}\\
& \\
\textbf{Planetary parameters} & \\
$M_p$ (M$_{J}$)                                 \dotfill & \hatcurPPmlong   & \hatcurPPmlongeccen & \boldmath{\hatcurPPmlongout}\\
$R_p$ (R$_{J}$)                                 \dotfill & \hatcurPPrlong   & \hatcurPPrlongeccen & \boldmath{\hatcurPPrlongout}\\
$C(M_p, R_p)$$^e$                                      \dotfill & \hatcurPPmrcorr &  \hatcurPPmrcorreccen & \boldmath{\hatcurPPmrcorrout}\\
$\rho_p$ (g/cm$^3$)                                    \dotfill & \hatcurPPrho     & \hatcurPPrhoeccen & \boldmath{\hatcurPPrhoout}\\
$\log g_p$ (cgs)                                       \dotfill & \hatcurPPlogg   &  \hatcurPPloggeccen & \boldmath{\hatcurPPloggout}\\
$a$ (AU)                                               \dotfill & \hatcurPParel   &  \hatcurPPareleccen & \boldmath{\hatcurPParelout}\\
$T_{eq}$ (K)                                           \dotfill & \hatcurPPteff    & \hatcurPPteffeccen & \boldmath{\hatcurPPteffout}\\
$\Theta$$^f$                                         \dotfill & \hatcurPPtheta   & \hatcurPPthetaeccen & \boldmath{\hatcurPPthetaout}\\
$\langle F \rangle$ (10$^8$erg s$^{-1}$ cm$^{-2}$)$^g$ \dotfill & \hatcurPPfluxavg & \hatcurPPfluxavgeccen & \boldmath{\hatcurPPfluxavgout}\\ %
\hline
\end{tabular}
\tablefoot{\begin{description}
\item[$^a$]{$T_C$: Reference epoch of mid transit that minimizes the
correlation with the orbital period. BJD is calculated from UTC.
$T_{14}$: total transit duration, time between first to last
contact; $T_{12} = T_{34}$: ingress/egress time, time between
first and second, or third and fourth, contact.}
\item[$^b$]{Reciprocal of the half duration of the transit used as a jump
parameter in our MCMC analysis in place of $a/R_{\star}$. It is
related to $a/R_{\star}$ by the expression $\zeta/R_{\star} =
a/R_{\star}\cdot(2\pi(1+e\sin
\omega))/(P\sqrt{1-b^2}\sqrt{1-e^2})$ \citep{bakos2010}.}
\item[$^c$]{Values for a quadratic law given separately for the Sloan $g$,
$r$, $i$ and $z$ filters. These values were adopted from the
tabulations by \citet{claret2004} according to the spectroscopic
(SME) parameters listed in Table~\ref{tab:stellar_parameter}.}
\item[$^d$]{the uncertainties on the eccentricity $e$ incorporate the estimated RV jitter}
\item[$^e$]{Correlation coefficient between the planetary mass $M_p$ and 
radius $R_p$.}
\item[$^f$]{The Safronov number is given by $\Theta = (1/2)(V_{esc}/V_{orb})^2
= (a/R_p)\cdot(M_p/M_{\star})$ (see \citealp{hansen2007}).}
\item[$^g$]{Incoming flux per unit surface area, averaged over the orbit.}
\end{description}
}
\end{table*}

%
\section{Starspot analysis} %
\begin{figure*}
\centering
\includegraphics[trim = 0mm 0mm 30mm 130mm, clip,width=16.0cm]{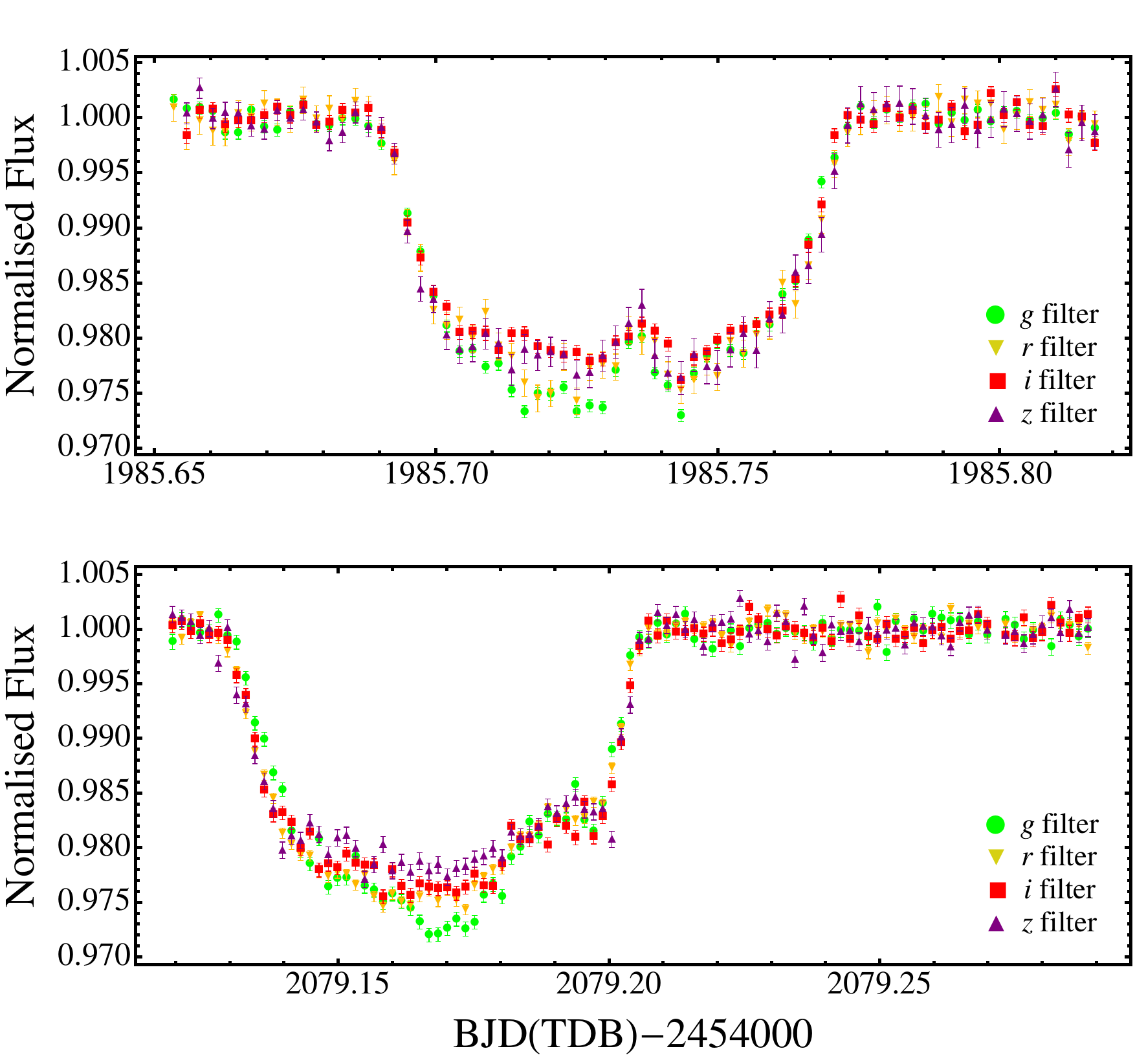}
\caption{
Combined four-colour transit light curves of HATS-2 obtained with the
GROND imaging system.  Green dots are for the data taken in the $g$
band, yellow upside down triangles for the $r$ band, red squares for
the $i$ band, and purple triangles for the $z$ band.  \emph{Top panel:}
transit observed on February 28, 2012.  The \emph{bump} observed just
after the midtransit is interpreted as the covering of a ``cold''
starspot by the planet.  \emph{Lower panel:} transit observed on June
1, 2012.  In addition to the bump occurred near the egress part of the
light curve, a ``hot'' spot manifested in the $g$ band, just before the
starting of the covering of the starspot.
}
\label{fig:grond-lc}
\end{figure*}

\label{sec:4}
Fig.~\ref{fig:grond-lc} shows the combined four-colour GROND light
curves for the two HATS-2 transit events that were observed with this
imaging instrument.  The slight difference in the transit depth among
the datasets is due to the different wavelength range covered by each
filter.  In particular, the $g$, $r$, $i$ and $z$ filters are sensitive
to wavelength ranges of $3860-5340$ \AA, $5380-7060$ \AA,
$7160-8150$ \AA, and $8260-9520$ \AA, respectively.
%
%

%
\subsection{Starspots and plages} %
\label{sec:4.1}
From an inspection of Fig.~\ref{fig:grond-lc}, it is easy to note
several distortions in the light curves.  Such anomalies cannot be
removed by choosing different comparison stars for the differential
photometry, and we interpret them as the consequence of the planet
crossing irregularities on the stellar photosphere, i.e.~starspots.  It
is well known that starspots are at a lower temperature than the rest
of the photosphere.  The flux ratio should be therefore lower in the
blue than the red.  We thus expect to see stronger starspot features in
the bluest bands.

The data taken on February 28, 2012, are plotted in the top panel of
Fig.~\ref{fig:grond-lc}, where the \emph{bump}, which is clearly
present just after midtransit in all four optical bands, is explained
by a starspot covered by the planet.  In particular, considering the
errorbars, the $g$, $r$, and $i$ points in the starspot feature look as
expected, whereas the feature in $z$ is a bit peaked, especially the
highest points at the peak of the bump at roughly BJD(TDB) 2455985.735. 
Before the starspot feature, it is also possible to note that the
fluxes measured in the $g$ and $r$ bands are lower than those in the
other two reddest bands, as if the planet were occulting a hotter zone
of the stellar chromosphere.  Actually, the most sensitive indicators
of the chromospheric activity of a star in the visible spectrum are the
emission lines of Ca\,II H$\lambda$3968, K$\lambda$3933, and H$\alpha$
$\lambda$6563, which in our case fall on the transmission wings of the
$g$ and $r$ GROND passbands. The characterization of the chromospheric activity by calculating the Ca activity indicator using FEROS spectra was not possible due to high noise in the spectra.

Within the transit observed on June 1, 2012, whose data are plotted in
the lower panel of Fig.~\ref{fig:grond-lc}, we detected another
starspot, which occurred near the transit-egress zone of the light
curve.  Again, before the starspot feature, we note another ``hotspot''
in the $g$ band, which has its peak at roughly BJD(TDB) 2456079.681.

These hotspot distortions could be caused by differential color
extinction or other time-correlated errors (i.e.~red noise) of
atmospheric origin. The g-band suffers most from the strength and
  variability of Earth-atmospheric extinction of all optical wavelengths
  covered by GROND, why the distortions in the g-band could have an
  atmospheric origin. Discrepancies in blue filters have been noted by other observers, and are often ascribed to systematic errors in ground-based photometry with these filters (e.g. \citealp{southworth2012}). However, our group has observed more than 25 planetary transits with the GROND instrument to date, and in no other case have we seen similar features in the $g$-band only. We consider it unlikely that a systematic error of this form would only appear near to other spot features in the HATS-2 light curve, and therefore conclude that a more plausible scenario is that of a ``plage''.  A plage is a chromospheric region typically located near
active starspots, and usually forming before the starspots appear, and
disappearing after the starspots vanish from a particular area (e.g. \citealp{Carroll1996}). 
Accordingly, a plage occurs most often near a starspot region.  As a
matter of fact, in the GROND light curves, our plages are located just
before each starspot.  One can argue that the plage in the second
transit is visible only in the $g$ band, but this can be explained by
temperature fluctuations in the chromosphere, which causes a lack of
ionized hydrogen, and by the fact that the Ca\,II lines are much
stronger than the H$\alpha$ line for a K-type star like HATS-2. 
Another argument supporting this \emph{plage--starspot} scenario is
that, for these old stars, a solar-like relation between photospheric
and chromospheric cycles is expected, the photospheric brightness
varying in phase with that of the chromosphere \citep{lockwood2007}.

We note, however, that if these are plages they must be rather
different from solar plages, which are essentially invisible in
broad-band optical filters unless they are very close to the solar
limb. Detecting a plage feature through a broad-band filter near the
stellar center suggests a much larger temperature contrast and/or
column density of chromospheric gas than in the solar case.
%
\subsection{Modelling transits and starspots} %
\label{sec:4.2}

We modelled the GROND transit light curves of HATS-2 with the
PRISM\footnote{PRISM (Planetary Retrospective Integrated Star-spot
Model).} and GEMC codes \citep{tregloan2013}.  The first code models a
planetary transit over a spotted star, while the latter one is an
optimisation algorithm for finding the global best fit and associated
uncertainties.  Using these codes, one can determine, besides the ratio
of the radii $R_{p}/R_{\star}$, the sum of the fractional radii,
$r_{p}+r_{\star} = (R_{p}+R_{\star})/a$, the limb darkening
coefficients, the transit midpoint $T_{0}$, and the orbital inclination
$i_p$, as well as the photometric parameters of the spots, i.e.~the
projected longitude and the latitude of their centres ($\theta$ and
$\phi$, these are equal to the physical latitude and longitude only if the rotation axis of the star is perpendicular to the line of sight), the spot size $r_{\mathrm{spot}}$ and the spot contrast
$\rho_{\mathrm{spot}}$, which is basically the ratio of the surface
brightness of the spot to that of the surrounding photosphere. 
Unfortunately, the current versions of PRISM and GEMC are set to
fit only a single starspot (or hotspot), so we excluded the $g$-band
dataset of the $2^{\mathrm{nd}}$ transit from the analysis, because it
contains a hotspot with high contrast ratio between stellar photosphere
and spot, which strongly interferes with the best-fitting model for the
light curve.

Given that the codes do not allow the datasets to be fitted
simultaneously, we proceeded as follows.  First, we modelled the seven
datasets ($1^{\mathrm{st}}$ transit: $g$, $r$, $i$, $z$;
$2^{\mathrm{nd}}$ transit: $r$, $i$, $z$) of HATS-2 separately; this
step allowed us to restrict the search space for each parameter.  Then,
we combined the four light curves of the first transit into a single
dataset by taking the mean value at each point from the four bands at
that point and we fitted the corresponding light curve; this second
step was necessary to find a common value for $T_0$, $i_p$, $\theta$ and
$\phi$.  Finally, we fitted each light curve separately fixing the
starspot position, the midtime of transit $T_{0}$ and the system
inclination to the values found in the previous combined fit. While these parameters are the same for each band since they are physical parameters of the spot or the system and are therefore fixed during the analysis, other parameters as radius of the
planet $R_p$, spot contrast $\rho_{spot}$ and temperature of the starspots $T_{spot}$ change according to the wavelength and hence according to the analysed band and are therefore free parameters during the fit.

\begin{figure*}
\centering
\includegraphics[trim = 0mm 0mm 0mm 120mm, clip,width=16.0cm]{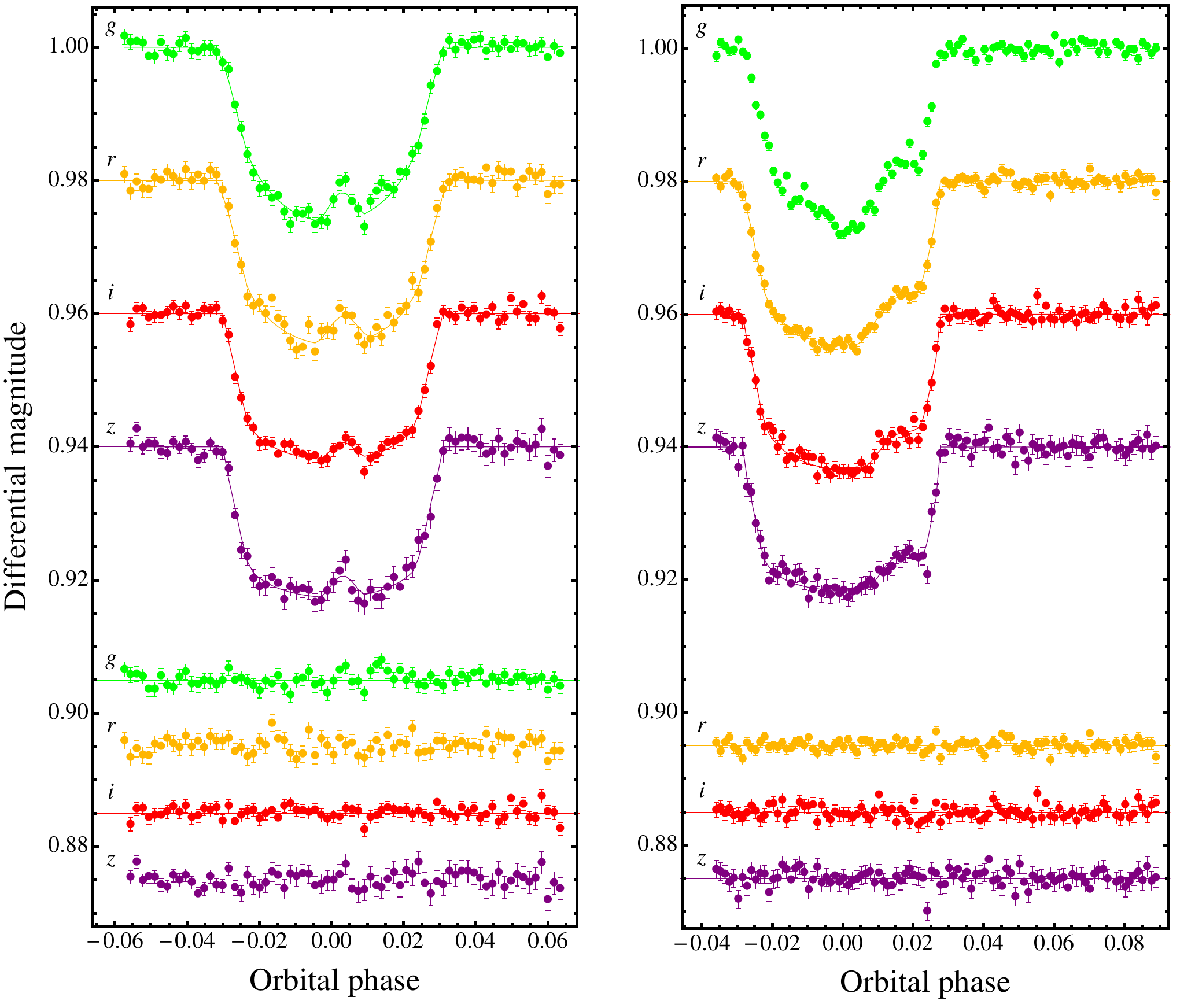}
\caption{
Phased GROND light curves of HATS-2b compared to the best GEMC fits. 
The light curves and the residuals are ordered according to the central
wavelength of the filter used.  \emph{Left panel:} transit observed on
February 28, 2012.  \emph{Right panel:} transit observed on June 1,
2012; due to to presence of the hotspot, the $g$ band was not analysed
with GEMC (see text).
}
\label{fig:best-fitting-lc}
\end{figure*}

The light curves and their best-fitting models are shown in
Fig.~\ref{fig:best-fitting-lc}, while the derived photometric
parameters for each light curve are reported in
Table~\ref{tab:photometric-parameters-1}, together with the results of
the MCMC error analysis for each solution.

\begin{table*}
\centering %
\caption{
Photometric parameters derived from the GEMC fitting of the GROND
transit light curves.
}
\label{tab:photometric-parameters-1}
\begin{tabular}{l c | c c c c}%
\hline %
\hline %
$1^{\mathrm{st}} \, transit$ & & & & & \\
\hline %
Parameter & Symbol & $g$ & $r$ & $i$ & $z$ \\
\hline %
Radius ratio                  & $R_{p}/R_{\star}$             & $0.1348 \pm 0.0011$ & $0.1324 \pm 0.0011$ & $0.13145\pm 0.00096$ & $0.1352 \pm 0.0010$ \\
Sum of fractional radii       & $r_{p}+r_{\star}$             & $0.2204 \pm 0.0018$ & $0.2232 \pm 0.0018$ & $0.2149 \pm 0.0016 $ & $0.2161 \pm 0.0015$ \\
Linear LD coefficient         & $u_{1}$                       & $0.749  \pm 0.060 $ & $0.593  \pm 0.051 $ & $0.352  \pm 0.057  $ & $0.298  \pm 0.039 $ \\
Quadratic LD coefficient      & $u_{2}$                       & $0.171  \pm 0.018 $ & $0.296  \pm 0.043 $ & $0.218  \pm 0.028  $ & $0.144  \pm 0.024 $ \\
Inclination (degrees) $^{a}$         & $i_p$                    & $85.26  \pm 0.40$   & $85.26  \pm 0.40$   & $85.26  \pm 0.40$    & $85.26  \pm 0.40$   \\
Longitude of spot (degrees) $^{a,b}$ & $\theta$               & $5.78   \pm 0.45$   & $5.78   \pm 0.45$   & $5.78   \pm 0.45$    & $5.78   \pm 0.45$   \\
Latitude of Spot (degrees) $^{a,c}$  & $\phi$                 & $76.52  \pm 1.94$   & $76.52  \pm 1.94$   & $76.52  \pm 1.94$    & $76.52  \pm 1.94$   \\
Spot angular radius (degrees) $^{d}$ & $r_{\mathrm{spot}}$    & $8.85   \pm 0.61 $  & $10.01  \pm 0.72  $ & $8.93   \pm 0.65   $ & $8.72   \pm 0.50  $ \\
Spot contrast $^{e}$                 & $\rho_{\mathrm{spot}}$ & $0.304  \pm 0.035$  & $0.546  \pm 0.048 $ & $0.464  \pm 0.052  $ & $0.251  \pm 0.52  $ \\
\hline %
\hline %

$2^{\mathrm{nd}} \, transit$ & & & & & \\
\hline %
Radius ratio                  & $R_{p}/R_{\star}$             & $-$ & $0.1356 \pm 0.0012$ & $0.13411\pm 0.00093$ & $0.1307 \pm 0.0011 $ \\
Sum of fractional radii       & $r_{p}+r_{\star}$             & $-$ & $0.2108 \pm 0.0019$ & $0.2022 \pm 0.0012$  & $0.2039 \pm 0.0017 $ \\
Linear LD coefficient         & $u_{1}$                       & $-$ & $0.473  \pm 0.057 $ & $0.399  \pm 0.049 $  & $0.252  \pm 0.044  $ \\
Quadratic LD coefficient      & $u_{2}$                       & $-$ & $0.250  \pm 0.038 $ & $0.230  \pm 0.025 $  & $0.316  \pm 0.049  $ \\
Inclination (degrees) $^a$           & $i_p$                    & $-$ & $85.89  \pm 0.40$   & $85.89  \pm 0.40$    & $85.89  \pm 0.40$    \\
Longitude of spot (degrees) $^{a,b}$ & $\theta$               & $-$ & $35.26  \pm 1.20$   & $35.26  \pm 1.20$    & $35.26  \pm 1.20$    \\
Latitude of Spot (degrees)  $^{a,c}$ & $\phi$                 & $-$ & $80.60  \pm 2.10$   & $80.60  \pm 2.10$    & $80.60  \pm 2.10$    \\
Spot angular radius (degrees) $^{d}$ & $r_{\mathrm{spot}}$    & $-$ & $20.14  \pm 1.49  $ & $17.79  \pm 2.17  $  & $18.28  \pm 2.31   $ \\
Spot contrast $^{e}$                 & $\rho_{\mathrm{spot}}$ & $-$ & $0.753  \pm 0.046 $ & $0.780  \pm 0.054 $  & $0.789  \pm 0.047  $ \\
\hline %
\hline
\end{tabular}
\tablefoot{\begin{description}
\item[$^a$]{This is a common value and was found from the preceding fit of the combined data (see text).}
\item[$^b$]{The longitude of the centre of the spot is defined to be $0^{\circ}$ at the centre of the stellar disc and can vary from $-90^{\circ}$ to $90^{\circ}$.}
\item[$^c$]{The latitude of the centre of the spot is defined to be $0^{\circ}$ at the north pole and $180^{\circ}$ at the south pole.}
\item[$^d$]{Note that $90^{\circ}$ degrees covers half of stellar surface.}
\item[$^e$]{Note that 1.0 equals to the surrounding photosphere.}
\end{description}}
\end{table*}
%

Comparing Table~\ref{tab:orbit_parameter} with
Table~\ref{tab:photometric-parameters-1} we find that the fitted light
curve parameters from the analysis described in Section
\ref{sec:phot_rv_analysis} are consistent with the parameters that
result from the GEMC+PRISM model, except for the inclination which
differs by more than $2\sigma$.  As already discussed in Section
\ref{sec:phot_rv_analysis}, the joint-fit analysis was performed both
considering and without considering the points contaminated by the
starspots, and the results are consistent with each other.  So, our
conclusion is that the spots themselves are not systematically
affecting the stellar or planet parameters in a significant way; the
differences in the inclination between GEMC and our joint fit are most
likely due to differences in the modeling.

%

%
\subsection{Starspots discussion} %
\label{sec:4.4}
The final value for the starspots angular radii comes from the weighted
mean of the results in each band and is $r_{\mathrm{spot}}=9.02^{\circ}
\pm 0.30^{\circ}$ for the starspot in the $1^{\mathrm{st}}$ transit
(spot \#1) and $r_{\mathrm{spot}}=19.16^{\circ} \pm 1.08^{\circ}$ for
the starspot in the $2^{\mathrm{nd}}$ transit (spot \#2), with a
reduced $\chi_{\nu}^2$ of 0.78 and 0.49 respectively, indicating a good agreement between the various light curves in each of the
two transits.  We note that the error of the angular size of the spot
\#2 is greater than that of the spot \#1.  While it may be that spot
\#2 is larger, we caution that its position near the limb of the star
makes it size poorly constrained.

The above numbers translate to radii of $98\,325 \pm 3\,876$\,km and
$208\,856 \pm 11\,794$\,km, which are equivalent $2.5\%$ and
$11\%$ of the stellar disk, respectively.  Starspot sizes are in
general estimated by doppler-imaging reconstructions
(i.e.~\citealp{cameron1992,vogt1999}) and their range is $0.1\%$ to
$22\%$ of a stellar hemisphere, the inferior value being the detection
limit of this technique \citep{strassmeier2009}.  Our measurements are
thus perfectly reasonable for a common starspot or for a starspot
\emph{assembly}, and in agreement with what has been found in other
K-type stars (e.g. TrES-1 (a K0V star) reveals a starspot of at least 42\,000\,km in radius, see \citet{rabus2009}). 

Starspots are also interesting in terms of how the contrast changes
with passband.  In particular, we expect that moving from ultraviolet
(UV) to infrared (IR) wavelengths the spot becomes brighter relative to the photosphere. 
Considering the starspot \#2, its contrast decreases from $r$ to $z$,
even though this variation is inside the 1$\sigma$ error (see Table~\ref{tab:photometric-parameters-1}).  Considering that HATS-2 has an
effective temperature $T_{eff \star}= 5227 \pm 95$ K and modeling both
the photosphere and the starspot as blackbodies
\citep{rabus2009,sanchis2011}, we used Eq.  (1) of \citet{silva2003} to
estimate the temperature of the starspot \#2 in each band:
\begin{equation}
f_i = \frac{\exp (h\nu/k_BT_e) - 1}{\exp(h\nu/k_BT_0) - 1}
\end{equation}
with the spot contranst $f_i$, the Planck constant $h$, the frequency of the observation $\nu$, the effective surface temperature of the star $T_e$ and the spot temperature $T_0$. We obtained the following values: $T_{\mathrm{spot}\#2,r}= 4916 \pm 105$ K,
$T_{\mathrm{spot}\#2,i}= 4895 \pm 121$ K and $T_{\mathrm{spot}\#2,z}=
4856 \pm 120$ K.  The weighted mean is $T_{\mathrm{spot}\#2}=4891.5 \pm
66.2$ K.

\begin{figure}
\centering
\includegraphics[trim = 0mm 0mm 45mm 190mm, clip,width=9.0cm]{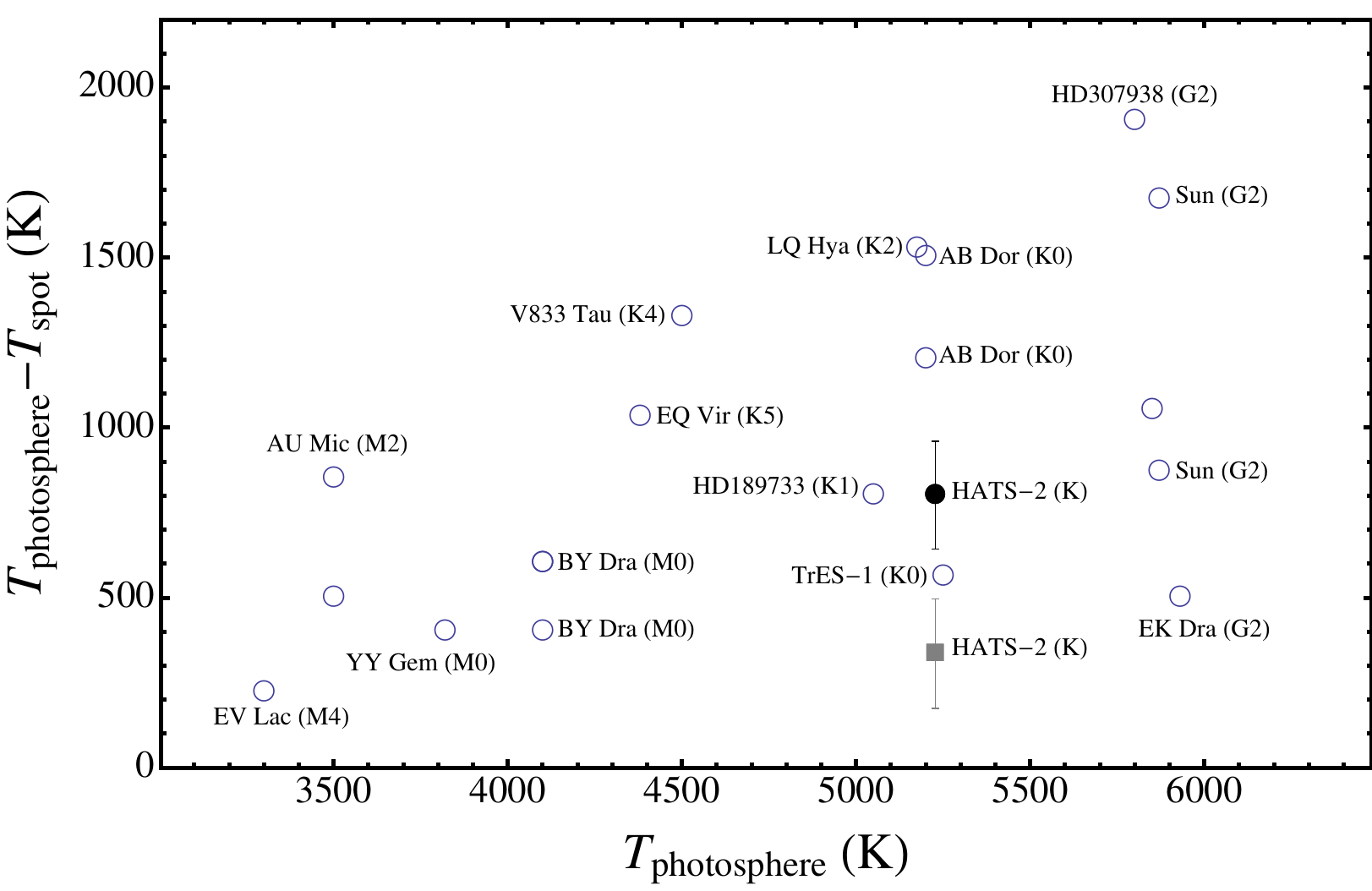}
\caption{
Spot temperature contrast with respect to the photospheric temperature
in several dwarf stars.  Gray square is from spot \#2, black circle
from spot \#1, open circles from \citet{berdyugina2005}, except TrES-1 \citep{rabus2009} and HD189733 \citep{sing2011}.  The name of
the star and its spectral type is also reported for most of them. Nameless targets do not have a name in the publication of \citet{berdyugina2005} as well. Note
that some stars appear two times.
}
\label{fig:spot-contrast}
\end{figure}

Unlike starspot \#2, the spot contrasts for starspot \#1 are
inconsistent with expectations.  The spot is too bright in $r$ relative
to the other bandpasses, and too faint in $z$.  If we estimate the
starspot temperature in each band, we find $T_{\mathrm{spot}\#1,g}=
4345 \pm 97$ K, $T_{\mathrm{spot}\#1,r}= 4604 \pm 109$ K,
$T_{\mathrm{spot}\#1,i}= 4318 \pm 128$ K and $T_{\mathrm{spot}\#1,z}=
3595 \pm 180$ K.  While the temperature in $r$ is in agreement with
those of $g$ and $i$ at $1-2\sigma$ level, and slight differences could be
explained by chromospheric contamination (filaments, spicules, etc.), the
temperature in $z$ seems physically inexplicable.  This effect is
essentially caused by the $z$ points at the peak of the starspot, at
phase $\sim 0.004$ (see Fig.~\ref{fig:best-fitting-lc}), which are
higher than the other points.  However, one has also to consider that
errorbars in this band are larger than those found in the other bands. 
This is due to the fact that, since the GROND system design does not
permit to chose different exposure times for each band, we are forced
to optimize the observations for the $r$ and $i$ bands.  Consequently,
considering both the filter-transmission efficiency and the color and
the magnitude of HATS-2, the SNR in these two bands is better than that
in $z$, for which we have larger uncertainty in the photometry.
Taking these considerations into account, we estimated the final
temperature of starspot \#1 neglecting the $z$-band value, and
obtaining $T_{\mathrm{spot}\#1}=4425 \pm 63$ K.  In
Fig.~\ref{fig:spot-contrast} the final values of the temperature
contrast of the two starspots are compared with those of a sample of
dwarf stars, which was reported by \citet{berdyugina2005}.  The derived
contrast for the HATS-2 starspots is consistent with what is observed
for other stars.
%
As already observed by \citet{strassmeier2009}, the temperature
difference between photosphere and starspots can be not so different
for stars of different spectral types.  Moreover, in the case of long
lifetime, the same starspot could been seen at quite different
temperature \citep{kang1989}.  It is then very difficult to find any
clear correlation between starspot temperatures and spectral classes of
stars.

The achieved longitudes of the starspots are in agreement with a visual
inspection of the light curves.  The latitude of starspot \#1,
$76.52^{\circ} \pm 1.94^{\circ}$, matches well with that of starspot
\#2, $80.6^{\circ} \pm 2.1^{\circ}$, the difference being within
1$\sigma$.  

Multiple planetary transits across the same spot complex can be used
to constrain the alignment between the orbital axis of the planet and
the spin axis of the star (e.g. \citealp{sanchis2011b}).
Unfortunately, from only two transits separated by 94 days we cannot
tell whether or not the observed anomalies are due to the same
complex. It is possible that they are. Following \citet{solanki2003} we
estimate a typical lifetime of $\sim 130$ days for spots of the size
seen here. Moreover the rotation period of $P_{\rm rot} = 31 \pm 1$\,d
inferred assuming they are the same spot is consistent with the value
of $P_{\rm rot} = (30 \pm 10~{\rm d}) \sin i_{\star}$ estimated from
the spectroscopically-determined sky-projected equatorial rotation
speed. If they are the same spot complex, then the sky-projected
spin-orbit alignment is $\lambda = 8^{\rm \circ} \pm 8^{\rm \circ}$,
which is consistent with zero. We caution, however, that this value
depends entirely on this assumption which could easily be wrong.
Continued photometric monitoring of HATS-2, or spectroscopic
observations of the Rossiter-McLaughlin effect, are necessary to
measure the spin-orbit alignment of this system.\\
\\
To test whether the spot parameters inferred from modelling the transits are consistent with the amplitude of variations seen in the HATSouth photometry, we simulate a light curve using the Macula starspot model \citep{kipping:2012:macula} and the model parameters determined from the first GROND $r$-band transit. We find that such a spot gives rise to periodic variations with a peak-to-peak amplitude of $\sim 5$\,mmag, which is within the 3.6 to 10.0\,mmag range of amplitudes observed in the HATSouth light curve. The fact that the amplitude changes by a factor of $\sim 3$ over the course of the HATSouth observations indicates, however, that the spot(s) observed by HATSouth is(are) likely to be unrelated to the spot(s) observed with GROND.

%
\section{Conclusions}
\label{sec:5}
\begin{figure*}
\centering
\includegraphics[trim = 20mm 125mm 90mm 00mm, clip,width=15.6cm]{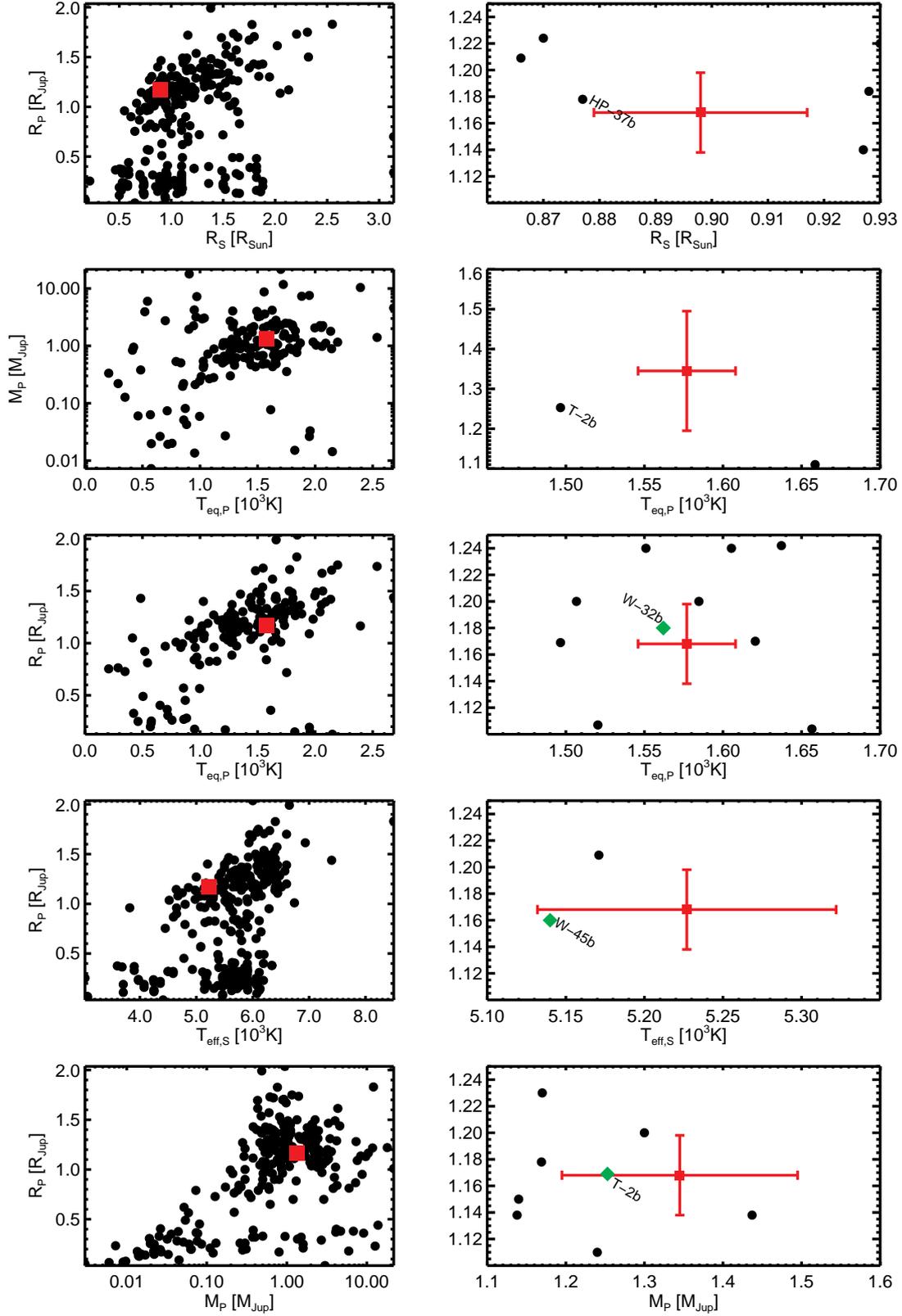}
\caption{
Correlation diagrams for confirmed transit planets (exoplanet.eu, last
updated Jan 10 2013).  From top to bottom: planetary radius $R_p$
vs.~stellar radius $R_{\star}$, planetary mass $M_P$ vs.~planetary
equilibrium temperature $T_{eq,P}$, planetary radius $R_p$
vs.~planetary equilibrium temperature $T_{eq,P}$, planetary radius
$R_p$ vs.~stellar effective temperature $T_{eff,\star}$ and planetary
radius $R_p$ vs.  planetary mass $M_P$.  The position of the HATS-2 and
HATS-2b parameters, respectively, are marked in red squares.  Left
panels give a global overview of the position of HATS-2 and HATS-2b in
the population of known transit planets (black filled circles).  Right
panels illustrate a zoom in; green diamonds represent known exoplanets
which fall within the errorbars of the HATS-2 system parameters.  Short cuts:
HP = HAT-P, W = WASP, T = TrES.
}
\label{fig:correlation}
\end{figure*}

In this paper we have presented HATS-2b, the second planet discovered
by the HATSouth survey. This survey is a global network of six
identical telescopes located at three different sites in the Southern
hemisphere \citep{bakos2012b}.  The parameters of the planetary system
were estimated by an accurate joint fit of follow-up RV and photometric
measurements.  In particular, we found that HATS-2b has a mass of
\hatcurPPmlong$M_{J}$, and a radius of \hatcurPPrlong$R_{J}$.  To set
this target in the context of other transit planet detections, we
plotted 4 different types of correlation diagrams for the population of
transiting planets (Fig.~\ref{fig:correlation}).  We analysed the
location of determined parameters for HATS-2b and its host star HATS-2
in the following parameter spaces: planetary radius $R_p$ vs.~stellar
radius $R_{\star}$, planetary mass $M_P$ vs.~planetary equilibrium
temperature $T_{eq,P}$, planetary radius $R_p$ vs.~planetary
equilibrium temperature $T_{eq,P}$, planetary radius $R_p$ vs.~stellar
effective temperature $T_{eff,\star}$, and planetary radius $R_p$
vs.~planetary mass $M_P$.  As illustrated in
Fig.~\ref{fig:correlation}, the analysed parameter relations lie well
within the global distribution of known exoplanets.

Within each correlation diagram, at least one well characterized
exoplanet can be found, whose parameters are consistent with those of
the HATS-2 system within the error bars.  Looking on the correlation between
planetary and stellar radius, the HATS-2 system is almost like the HAT-P-37 system \citep{bakos2012a}. 
Comparing the planetary equlibrium temperature and planetary mass,
HATS-2b is similar to TrES-2b \citep{odonovan2006}.  The relation between planetary
equilibrium temperature and planetary mass shows an agreement with
WASP-32b \citep{maxted2010}, while the relation between stellar effective temperature and
planetary radius points out that HATS-2b agrees well with WASP-45b \citep{anderson2012}
within the error bars.  The focus on the planetary parameters radius
and mass reveals a similarity to the transiting planet TrES-2. 
Comparing the atmospheres of exoplanets with similar physical
parameters will be especially important to pursue with e.g.~the future
ECHO space mission \citep{Tinetti:2012}.
%

Very interesting is the detection of anomalies in the two multi-band
photometric-follow-up light curves obtained with the GROND imaging
instrument.  We recognize the anomalies as starspots covered by HATS-2b
during the two transit events, and used PRISM and GEMC codes
\citep{tregloan2013} to re-fit the transit light curves, measuring the
parameters of the spots.  Both the starspots appear to have associated
hot-spots, which appeared in the 1$^{\mathrm{st}}$ transit in the $g$
and $r$ bands, and only in the $g$ band in the 2$^{\mathrm{nd}}$
transit.  These hotspots could be physically interpreted as
chromospheric active regions known as \emph{plages}, which can be seen only in
the GROND's bluest bands.  We estimated the size and the temperature of
the two starspots, finding values which are in agreement with those
found in other G-K dwarf stars.

%
\begin{acknowledgements}

Development of the HATSouth project was funded by NSF MRI grant
NSF/AST-0723074, operations are supported by NASA grant NNX09AB29G, and
follow-up observations receive partial support from grant
NSF/AST-1108686.  Data presented in this paper is based partly on
observations obtained with the HATSouth Station at the Las Campanas
Observatory of the Carnegie Institution of Washington.
This work is based on observations collected at the MPG/ESO 2.2m
Telescope located at the ESO Observatory in La Silla (Chile), under
programme IDs P087.A-9014(A), P088.A-9008(A), P089.A-9008(A),
089.A-9006(A) and Chilean time P087.C-0508(A).  Operations of this
telescope are jointly performed by the Max Planck Gesellschaft and the
European Southern Observatory.  GROND has been built by the high-energy
group of MPE in collaboration with the LSW Tautenburg and ESO, and is
operating as a PI-instrument at the MPG/ESO 2.2m telescope.  We thank
Timo Anguita and R\'egis Lachaume for their technical assistance during
the observations at the MPG/ESO 2.2\,m Telescope.
A.J.\ acknowledges support from Fondecyt project 1130857, Anillo ACT-086,
BASAL CATA PFB-06 and the Millenium Science Initiative, Chilean Ministry of
Economy (Nucleus P10-022-F).  V.S.\ acknowledges support form BASAL CATA PFB-06.  R.B.\
and N.E.\ acknowledge support from Fondecyt project 1095213.
N.N. acknowledges support from an STFC consolidated grant.
M.R. acknowledges support from FONDECYT postdoctoral fellowship N$^\circ$3120097.
L.M.~thanks Jeremy Tregloan-Reed for his help in using of the PRISM
and GEMC codes, and John Southworth and Valerio Bozza for useful
discussions.

This paper also uses observations obtained with facilities of the Las
Cumbres Observatory Global Telescope.
Work at the Australian National University is supported by ARC Laureate 
Fellowship Grant FL0992131.
We acknowledge the use of the AAVSO Photometric All-Sky Survey (APASS),
funded by the Robert Martin Ayers Sciences Fund, and the SIMBAD
database, operated at CDS, Strasbourg, France.
Work at UNSW has been supported by ARC Australian Professorial Fellowship grant DP0774000,
ARC LIEF grant LE0989347 and ARC Super Science Fellowships FS100100046.
\end{acknowledgements}



\bibliographystyle{aa} 

\begin{thebibliography}{}
%
\bibitem[Alonso et al.(2008)]{alonso2008}
Alonso, R., Barieri, M., Rabus, M. et al. 2008, \aap, 487, L5
\bibitem[Alsubai et al.(2011)]{alsubai2011} %
Alsubai, K.~A., Parley, N.~R., Bramich, D.~M., et al. 2011,
\mnras, 417, 709
%
\bibitem[Anderson et al.(2012)]{anderson2012}
Anderson, D.~R., Collier Cameron, A., Gillon, M. et al. 2012, \mnras, 422, 1988
\bibitem[Armitage \& Bonnell(2002)]{armitage2002} %
Armitage, P.~J. \& Bonnell, I.~A. 2002, \mnras, 330, L11
%
\bibitem[Bakos et al.(2010)]{bakos2010}%
Bakos, G.~\'A., Torres, G., P\'al, A., et al. 2010, \apj, 710,
1724
%
\bibitem[Bakos et al.(2012a)]{bakos2012a} %
Bakos, G.~\'{A}., Hartman, J.~D., Torres, G., et al. 2012, \aj,
144, 19
%
\bibitem[Bakos et al.(2013)]{bakos2012b} %
Bakos, G.~\'{A}., Csubry, Z., Oenev, K., et al. 2013, \pasp,
125, 0
%
\bibitem[Ballester et al.(2007)]{ballester2007} %
Ballester, G.~E., Sing, D.~K., \& Herbert, F. 2007, \nat, 445, 511
%
\bibitem[Barnes(2009)]{barnes2009} %
Barnes, J.~W. 2009, \apj, 705, 683
%
\bibitem[Barros et al.(2011)]{barros2011}  
Barros, S.~C.~C., Pollacco, D.~L., Gibson, N.~P., et al. 2004,
\mnras, 416, 2593
%
\bibitem[Batalha et al.(2012)]{batalha2012} %
Batalha, N.~M., Rowe J.~F., Bryson S.~T., et al. 2012, \apjs, 204, 24
%
\bibitem[Bean et al.(2010)]{bean2010} %
Bean, J.~L., Miller-Ricci Kempton, E. \& Homeier, D. 2010, \nat,
468, 669
%
\bibitem[Berdyugina(2005)]{berdyugina2005} %
Berdyugina, S. V. 2005, Living Rev. Solar Phys., 2, 8
%
\bibitem[Bertin \& Arnouts(1996)]{bertin:1996} %
Bertin, E. \& Anouts, S. 1996, \aaps, 117, 393
%
\bibitem[Bonomo \& Lanza(2012)]{bonomo2012} %
Bonomo, A.~S., \& Lanza, A.~F. 2012, \aap, 547, A37
%
\bibitem[Borucki et al.(2009)]{borucki2009} %
Borucki, W.~J., Koch, D., Jenkins, J., et al. 2009, Science, 325,
709
%
\bibitem[Borucki et al.(2011a)]{borucki2011a} %
Borucki, W.~J., Koch, D.~G., Basri, G., et al. 2011a, \apj, 728,
117
%
\bibitem[Borucki et al.(2011b)]{borucki2011b} %
Borucki, W.~J., Koch, D.~G., Basri, G., et al. 2011b, \apj, 736,
19
%
\bibitem[Bryan et al.(2012)]{bryan2012} %
Bryan, M.~L., Alsubai, K.~A., Latham, D.~W., et al. 2012, ApJ,
750, 84
%
\bibitem[Carroll \& Ostlie(1996)]{Carroll1996} %
Carroll, B.~W. \& Ostlie, D.~A. 1996, An Introduction to Modern Astrophysics, Institute for Mathematics and Its Applications
\bibitem[Cassan et al.(2012)]{cassan2012} %
Cassan, A., Kubas, D., Beaulieu, J.-P., et al. 2012, \nat, 481,
167
%
\bibitem[Claret(2004)]{claret2004} %
Claret, A. 2004, \aap, 428, 1001
%
\bibitem[Collier Cameron(1992)]{cameron1992} %
Collier Cameron, A. 1992, Surface inhomogeneities on late-type
stars, ed. P. B. Byrne, \& D. J. Mullan (Springer, Berlin), 33
%
\bibitem[Collier Cameron et al.(2010)]{cameron2010} %
Collier Cameron, A., Guenther, E., Smalley, B., Mcdonald, I. 2010,
\mnras, 407, 507
%
\bibitem[D'Angelo et al.(2011)]{dangelo2011} %
D'Angelo, G., Durisen, R.~H., Lissauer, J.~J. 2011, Exoplanets,
edited by S. Seager. (University of Arizona Press), p. 319
%
\bibitem[D\'{e}sert(2011)]{desert2011} %
D\'{e}sert, J.-M., Charbonneau, D., Demory, B.-O., et al. 2011,
\apjs, 197, 14
%
\bibitem[Dopita et al.(2007)]{dopita2007} %
Dopita, M., Hart, J., McGregor, P., et al. 2007,  \apss, 310, 255
%
\bibitem[Fortney et al.(2008)]{fortney2008} %
Fortney, J.~J., Lodders, K., Marley, M.~S., \& Freedman, R.~S.
2008, \apj, 678, 1419
%
\bibitem[Fressin et al.(2013)]{fressin2013} %
Fressin, F., Torres, G., Charbonneau, D., et al. 2013, to appear
in \apj, arXiv:1301.0842
%
\bibitem[Gaudi et al.(2007)]{gaudi2007} %
Gaudi, B.~S. \& Winn, J.~N. 2007, \apj, 655, 550
%
\bibitem[Greiner et al.(2008)]{greiner2008} %
Greiner, J., Bornemann, W., Clemens, C., et al. 2008, \pasp, 120,
405
%
\bibitem[Gurdemir et al.(2012)]{gurdemir2012} %
Gurdemir, L., Redfield, S., \& Cuntz, M. 2012, \pasp 29, 141
%
\bibitem[Hansen \& Barman (2007)]{hansen2007} %
Hansen, B.~M.~S., \& Barman, T. 2007, \apj, 671, 861
%
\bibitem[Hartman et al.(2012)]{hartman2012} %
Hartman, J.~D., Bakos, G.~\'{A}., B\'{e}ky, B., et al. 2012, AJ,
144, 139
%
\bibitem[Hartman et al.(2011)]{hartman2011} %
Hartman, J.~D., Bakos, G.~\'A., Torres, G., et al. 2011, \apj,
742, 59
%
\bibitem[Hellier et al.(2012)]{hellier2012} %
Hellier, C., Anderson, D.~R., Collier Cameron, A., et al. 2012,
\mnras, 426, 739
%
\bibitem[Hirano et al.(2012)]{hirano2012} %
Hirano, T., Narita, N., Sato, B. et al. 2012, \apjl, 759, L36
%
\bibitem[Howard et al.(2012)]{howard2012} %
Howard, A.~W., Marcy, G.~W., Bryson, S.~T., et al. 2012, ApJS,
201, 15
%
\bibitem[Hussain(2002)]{hussain2002} %
Hussain, G.~A.~J 2002, Astron. Nachr., 323, 349
%
\bibitem[Kang \& Wilson(1989)]{kang1989} %
Kang, Y.~W. \& Wilson, R.~E. 1989, \aj, 97, 848
%
\bibitem[Kaufer \& Pasquini(1998)]{kaufer1998} %
Kaufer, A., \& Pasquini, L. 1998, Proc. SPIE, 3355, 844
%
\bibitem[Kipping et al.(2009)]{kipping2009} %
Kipping, D.~M., Fossey, S.~J., Campanella, G. 2009, \mnras, 400,
398
%
\bibitem[Kipping (2012)]{kipping:2012:macula}
Kipping, D.~M. 2012, \mnras, 427, 2487
%
\bibitem[Knutson et al.(2007)]{knutson2007} %
Knutson, H.~A., Charbonneau, D., Allen, L.~E., et al. 2007, \nat,
447, 183
%
\bibitem[Leconte et al.(2011)]{leconte2011} %
Leconte, J., Lai, D., Chabrier, G. 2011, \aap, 528, A41
%
\bibitem[Li et al.(2010)]{li2010} %
Li, S.-L., Miller, N. Lin, D.~N.~C., Fortney, J.~J. 2010, \nat,
463, 1054
%
\bibitem[Liu et al.(2011)]{liu2011}
Liu, H., Zhou, J-L., Wang, S. 2011, \apj, 732, 66
%
\bibitem[Lockwood et al.(2007)]{lockwood2007}
Lockwood, G.~W., Skiff, B.~A., Henry, G.~W., et al. 2007, \apjs,
171, 260
%
\bibitem[Lomb(1976)]{lomb:1976}
Lomb, N.~R. 1976, \apss, 39, 447
%
\bibitem[Lubow \& Ida(2011)]{lubow2011}
Lubow, S.~H., \& Ida, S. 2011, Exoplanets, edited by S. Seager.
(University of Arizona Press), p. 347
%
\bibitem[Mancini et al.(2013a)]{mancini2013a} %
Mancini, L., Southworth, J., Ciceri, S., et al. 2013, \aap, 551,
A11
%
\bibitem[Mancini et al.(2013b)]{mancini2013b} %
Mancini, L., Nikolov, N., Southworth, J., et al. 2013, to appear
in \mnras, arXiv:1301.3005
%
\bibitem[Mandel \& Agol(2002)]{mandel:2002} %
Mandel, K. \& Agol, E. 2002, \apjl, 580, L171
%
\bibitem[Maxted et al.(2010)]{maxted2010}
Maxted, P.~F.~L., Anderson, D.~R., Collier Cameron, A. et al. 2010, \pasp, 122, 1465
\bibitem[Mayor \& Queloz(1995)]{mayor1995} %
Mayor, M. \& Queloz, D. 1995, \nat, 378, 355
%
\bibitem[Mayor et al.(2011)]{mayor2011} %
Mayor, M., Marmier, M., Lovi, C., et al. 2011, arXiv1109.2497
%
\bibitem[Meibom et al.(2011)]{meibom:2011} %
Meibom, A., Barnes, S.~A., Latham, D.~W. et al. 2011, \apjl, 733, L9
%
\bibitem[Mordasini et al.(2012a)]{mordasini2012a} %
Mordasini, C., Alibert, Y., Klahr, H., Henning, T. 2012, \aap,
547, A111
%
\bibitem[Mordasini et al.(2012b)]{mordasini2012b} %
Mordasini, C., Alibert, Y., Georgy, C., et al. 2012, \aap, 547,
A112
%
\bibitem[O'Donovan et al.(2006)]{odonovan2006}
O'Donovan, F.~T., Charbonneau, D., Mandushev, G. et al. 2006, \apjl, 651, L61
\bibitem[Orosz et al.(2012a)]{orosz2012a} %
Orosz, J.~A., Welsh, W.~F., Carter, J.~A. et al. 2012a, \apj, 758,
87
%
\bibitem[Orosz et al.(2012b)]{orosz2012b} %
Orosz, J.~A., Welsh, W.~F., Carter, J.~A. et al. 2012b, Science,
337, 1511

\bibitem[P\"{a}tzold et al.(2013)]{patzold 2013} %
P\"{a}tzold, M., Endl, M., Csizmadia, Sz., et al. 2012, 545, A6
%
\bibitem[Penev et al.(2013)]{penev2013} %
Penev, K., Bakos, G.~\'{A}., Bayliss, D., et al. 2013, AJ, 145, 5
%
\bibitem[Petrovay \& van Driel-Gesztelyi(1997)]{petrovay1997} %
Petrovay, K., \& van Driel-Gesztelyi, L. 1997, Sol. Phys., 176,
249
%
\bibitem[Pont et al.(2007)]{pont2007} %
Pont, F., Gilliland, R.~L., Moutou, C., et al. 2007, \aap, 476,
1347
%
\bibitem[Press et al.(1992)]{press:1992}
Press, W.~H., Teukolsky, S.~A., Vetterling, W.~T. and Flannery, B.~P. 1992, Cambridge: University Press, 2nd ed.
%
\bibitem[Queloz et al.(2000a)]{queloz2000a} %
Queloz, D., Eggenberger, A., Mayor, M., Perrier, C., Beuzit,
J.-L., Naef, D., Sivan, J.-P., Udry, S. 2000, \aap, 359, L13
%
\bibitem[Queloz et al.(2000b)]{queloz2000b} %
Queloz, D., Mayor, M., Weber, L. et al. 2000, \aap, 354, 99
%
\bibitem[Rabus et al.(2009)]{rabus2009} %
Rabus, M., Alonso, R., Belmonte, J.~A, et al. 2009, \aap, 494, 391
\bibitem[Rabus et al.(2009b)]{rabus2009b} %
Rabus, M., Deeg, H.~J., Alonso, R. et al. 2009, \aap, 508, 1011
%
\bibitem[Rouan et al.(2012)]{rouan2012} %
Rouan, D., Parviainen, H., Moutou, C., et al. 2012, \aap, 537, A54
%
\bibitem[Sanchis-Ojeda \& Winn(2011)]{sanchis2011}
Sanchis-Ojeda, R., Winn, J. N. 2011, \apj, 743, 61
%
\bibitem[Sanchis-Ojeda et al.(2011)]{sanchis2011b}
Sanchis-Ojeda, R., Winn, J. N., Holman, M. J., et al. 2011, \apj,
733, 127
%
\bibitem[Scandariato et al.(2013)]{scandariato2013}
Scandariato, G., Maggio, A., Lanza, A.~F., et al. 2013,
arXiv:1301.7748
%
\bibitem[Scargle(1982)]{scargle:1982}
Scargle, J.~D. 1982, \apj, 263, 835
%
\bibitem[Schwamb et al.(2012)]{schwamb2012} %
Schwamb, M.~E., Orosz, J.~A., Carter, J.~A. et al. 2012, submitted
to ApJ, arXiv:1210.3612
%
\bibitem[Seager \& Sasselov(2000)]{seager2000} %
Seager, S., \& Sasselov, D.~D. 2000, \apj, 537, 916
%
\bibitem[Silva(2003)]{silva2003} %
Silva, A.~V.~R. 2003, \apjl, 585, L147
%
\bibitem[Sing et al.(2009)]{sing2009} %
Sing, D.~K., D\'{e}sert, J.-M., Lecavelier des Etangs, A., et al.
2009, \aap, 505, 891
%
\bibitem[Sing et al.(2011)]{sing2011}
Sing, D.~K., Pont, F., Aigrain, S. et al. 2011, \mnras, 416, 1443
\bibitem[Siverd et al.(2012)]{siverd2012} %
Siverd, R.~J., Beatty, T.~G., Pepper, J., et al. 2012, \apj, 761,
123
%
\bibitem[Shkolnik et al.(2008)]{shkolnik2008} %
Shkolnik, E., Bohlender, D.~A., Walker, G.~A.~H., Collier Cameron,
A. 2008, \apj, 676, 628
%
\bibitem[Skumanich(1972)]{skumanich:1972}
Skumanich, A. 1972, \apj, 171, 565
%
\bibitem[Smalley et al.(2012)]{smalley2012} %
Smalley, B., Anderson, D.~R., Collier-Cameron, A., et al. 2012,
\aap, 547, A61
%
\bibitem[Solanki(2003)]{solanki2003}
Solanki, S.~K. 2003, A\&ARv, 11, 153
%
\bibitem[Southworth et al.(2009)]{southworth2009}
Southworth, J., Hinse, T. C., J{\o}rgensen, U. G., et al., 2009,
\mnras, 396, 1023
%
\bibitem[Southworth et al.(2011)]{southworth2011} %
Southworth, J., Dominik, M., J{\o}rgensen, U. G., et al., 2011,
\aap, 527, A8
%
\bibitem[Southworth et al.(2012)]{southworth2012} %
Southworth, J., Mancini, L., Maxted, P.~F.~L., et al. 2012,
\mnras, 422, 3099
%
\bibitem[Steffen et al.(2013)]{steffen2013} %
Steffen, J.~H., Fabrycky, D.~C., Agol, E., et al. 2013, \mnras,
428, 1077
%
\bibitem[Strassmeier(2009)]{strassmeier2009}
Strassmeier, K. G. 2009, Astron. Astrophys. Rev., 17, 251
%
\bibitem[Swain et al.(2008)]{swain2008} %
Swain, M.~R., Vasisht, G., \& Tinetti, G. 2008, \nat, 452, 329
%
\bibitem[Szab\'{o} et al.(2011)]{szabo2011} %
Szab\'{o}, Gy. M., Szab\'{o}, R., Benk\~{o}, J.~M. et al. 2011,
\apj, 736, L4
%
\bibitem[Tinetti et al.(2012)]{Tinetti:2012} Tinetti, G., et al.~2012, 
Experimental Astronomy, 34, 311 
\bibitem[Tregloan-Reed et al.(2013)]{tregloan2013} %
Tregloan-Reed, J., Southworth, J., \& Tappert, C. 2013, \mnras,
428, 3671
%
\bibitem[Tusnski \& Valio(2011)]{tusnski2011}
Tusnski, L.~R.~M., \& Valio, A. 2011, \apj, 743, 97
%
\bibitem[Valenti \& Piskunov(1996)]{valenti1996} %
Valenti, J.~A. \& Piskunov, N. 1996, A\&AS, 118, 595
%
\bibitem[Vogt et al.(1999)]{vogt1999} %
Vogt, S.~S., Hatzes A., Misch A., K\"{u}rster, M. 1999, \apjs,
121, 547
%
\bibitem[Winn et al.(2010)]{winn2010} %
Winn, J.~N., Fabrycky, D., Albrecht, S., Johnson, J. A. 2010,
\apj, 718, L145
%
\bibitem[Yi et al.(2001)]{yi2001} %
Yi, S., Demarque, P., Kim. Y.-C., et al. 2001, \apjs, 136, 417
%
\bibitem[Alonso et al.(2004)]{alonso2004}%
Alonso, R., Brown, T.~M., Torres, G., et al. 2004, \apjl, 613, L153
%
\bibitem[McCullough et al.(2005)]{mccullough2005}%
McCullough, P.~R., Stys, J.~E., Valenti, J.~A., et al. 2005, \pasp, 117, 783
%
\bibitem[Zacharias et al.(2012)]{zacharias2012}%
Zacharias, N., Finch, C.~T., Girard, T.~M., et al. 2012, VizieR Online Data
Catalog, 1322, 0Z
\end{thebibliography}

\end{document}